\documentclass[a4paper,11pt]{article}
\pdfoutput=1 

\usepackage{jheppub} 

\usepackage[T1]{fontenc} 
\usepackage{graphicx}
\usepackage{dcolumn}
\usepackage{bm}
\usepackage{slashed}
\usepackage{commath}
\usepackage{multirow}
\usepackage{subcaption}
\usepackage{physics}
\usepackage{float}
\usepackage[dvipsnames]{xcolor}

\graphicspath{{figure2/}{figure1/}}

\title{New Limits on Coloured Three Jet Resonances}

\author{Hassan Easa, }
\author{Thomas Gregoire, and }
\author{Daniel Stolarski}
\affiliation{Ottawa-Carleton, Institute for Physics,\\
 Carleton University 1125 Colonel By Drive, \\
 Ottawa, ON, K1S 5B6, Canada}

\emailAdd{Hassaneasa@cmail.carleton.ca}
\emailAdd{gregoire@physics.carleton.ca}
\emailAdd{stolar@physics.carleton.ca}

\abstract{We consider experimental limits on colour triplet fermions that decay dominantly to three jets via a scalar mediator that can be on- or off-shell. These fermions arise in top-partner models that can solve the hierarchy problem, and limits on this scenario are weaker than those on traditional top-partner models because of the messy all-hadronic final state with significant backgrounds. We do find, however, that while there are no dedicated searches for this scenario, especially in case of an on-shell mediator, the suite of LHC all-hadronic searches still constrains a significant portion of the parameter space. In particular, we find that searches for pair production of di-jet and tri-jet resonances are complementary, covering different regions of parameter space. We also find that if the final state is rich in $b$-jets, current limits do not change significantly relative to the scenario with all light jets, and we describe how modifications of current search strategies can improve limits in that case.  }

\begin{document} 
\maketitle
\flushbottom

\section{\label{sec:level1}Introduction }

With the discovery of the Higgs boson~\cite{Aad:2012tfa, Chatrchyan:2012xdj} at the large hadron collider (LHC), the Standard Model of particle physics (SM) is complete. The confirmation of the properties of the Higgs being SM-like and the lack of discovery of new physics at the TeV scale exacerbates the hierarchy problem: what cuts off the quantum corrections to the Higgs mass? One well known solution to the hierarchy problem poses the existence of fermionic top partners, fermions with the same quantum numbers as the top quark whose contributions to the Higgs mass parameter cancel those of the top quark. These can appear in composite Higgs models~\cite{Contino:2006qr, Marzocca:2012zn, Matsedonskyi:2012ym, DeSimone:2012fs, Bellazzini:2014yua} and Little Higgs models~\cite{ArkaniHamed:2002qy, Schmaltz:2005ky}. 

In these models, top partners typically decay to a top quark and a Higgs or $Z$, or to a bottom quark and $W$. LHC searches for top partners in these modes are extensive, both in pair production~\cite{ Aaboud:2017qpr, Sirunyan:2017usq, Sirunyan:2017pks, Aaboud:2018xuw, Sirunyan:2018omb, Aaboud:2018uek, Aaboud:2018wxv, Aaboud:2018pii, Sirunyan:2018qau, Sirunyan:2019sza} and in single production~\cite{Aad:2014efa, Aad:2015voa, Aad:2016qpo, Khachatryan:2016vph, Sirunyan:2016ipo, Sirunyan:2017ezy, Sirunyan:2017tfc, Sirunyan:2017ynj, Sirunyan:2018fjh, Aaboud:2018saj, Sirunyan:2018ncp}, with limits $\approx $1.3--1.66 TeV on the mass of the top partners from the various searches depending on their branching ratios. Due to the lack of discovery, it is critical to explore alternative models, particularly those with different decay modes for the top partners. One could imagine, for example, decays involving a charged or neutral scalar~\cite{Anandakrishnan:2015yfa, Aguilar-Saavedra:2017giu, Aguilar-Saavedra:2019ghg}, decays involving a top quark in association with a gluon or a photon~\cite{Kim:2018mks, Alhazmi:2018whk},  a neutral boson that subsequently decays to two photons~\cite{Benbrik:2019zdp}, a dark photon or a dark Higgs~\cite{Kim:2019oyh}, or a pseudo-scalar which promptly decays to a pair of gluons or b-quarks~\cite{Chala:2017xgc, Bizot:2018tds, Cacciapaglia:2019zmj}. In this work we consider the particularly challenging possibility of a final state with only hadronic activity and no leptons or missing energy and study the limits on the masses of such top partners. 

\begin{figure}[tbp]
\centering 
\begin{subfigure}[b]{0.45\textwidth}
\includegraphics[width=.7\textwidth,origin=c,angle=0]{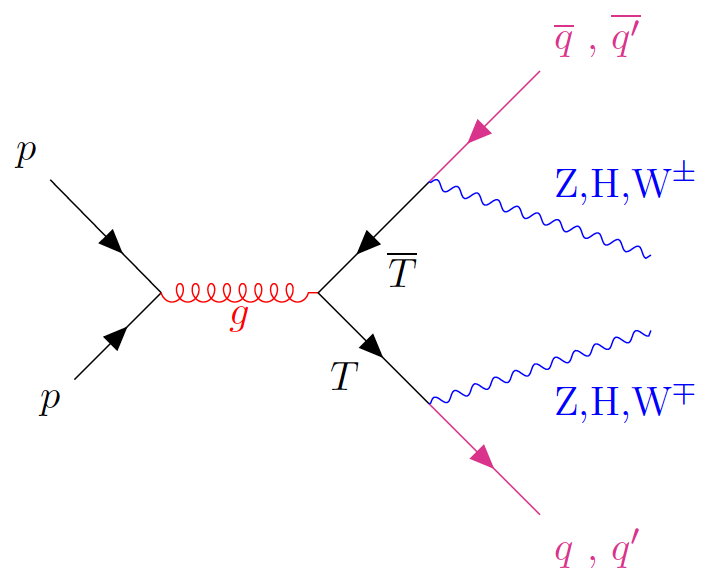}
\caption{Conventional decay modes}
\end{subfigure}
\quad
\begin{subfigure}[b]{0.45\textwidth}
\includegraphics[width=.7\textwidth,origin=c,angle=0]{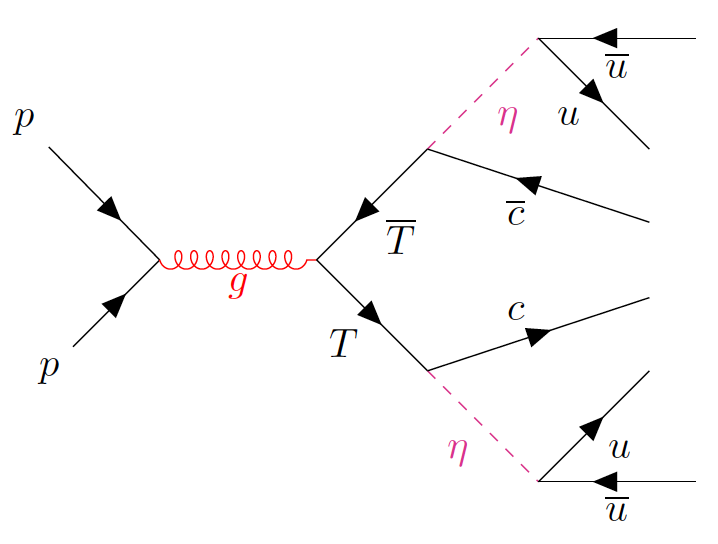}
\caption{Our model}
\end{subfigure}
\caption{\label{fig:feymanDiagrams_TopDecay} Feynman diagrams for the conventional decay modes and the processes considered in this letter.}
\end{figure}

We consider a simplified model for our top partners ($T$) which contains the following processes: $T$ is pair produced via the strong interactions, $pp \rightarrow T\bar{T}$. It then decays to a light flavour quark ($j$) and a new scalar $\eta$, and that scalar decays to two light flavour jets. The full process is
\begin{equation}
pp \rightarrow T\bar{T} \rightarrow jj\eta\eta \rightarrow 6j,
\label{eq:topology}
\end{equation}
with a representative Feynman diagram shown in the right panel of figure~\ref{fig:feymanDiagrams_TopDecay}. In our studies, we consider the decay channel $T \rightarrow j j j $ where $j$ can be associated with any light quark. However, if one were to focus mainly on the decay channel $T \rightarrow c j j $, the limits obtained from considering charm tagging will not significantly improve the sensitivity of the signal or alter the limits obtained, mainly because charm tagging is notoriously difficult and the efficiency is significantly worse than $b$-tagging. For example, the $c$-tagging efficiency is approximately 20$\%$ for about 1$\%$ light jet misidentification rate, where as the $b$-tagging efficiency can be approximately 70$\%$ at about 1$\%$ misidentification rate~\cite{TheATLAScollaboration:2015atd, CMS:2016knj, Sirunyan:2017ezt, Aaboud:2018xwy}. Note the tagging efficiencies depends on jet $p_{T}$ and $\eta$. An explicit model which gives rise to this signature (without associated top and bottom signatures) and solves the hierarchy problem is given in~\cite{Kats:2017ojr}. This scenario can be thought of as the fermionic analogue of hadronic $R$-parity violation (RPV)~\cite{Dreiner:1997uz, Allanach:2003eb, Barbier:2004ez} in Supersymmetry where the top squark can decay to two light jets.

This six-jet final state is experimentally very challenging as the QCD multijet background  is very large and difficult to determine. While one might expect that the limits for this model are significantly weaker than for the traditional decay modes, we will show that these models are also strongly constrained, with the best limits coming from recasting searches for RPV gluino searches from CMS~\cite{Sirunyan:2018duw} and ATLAS~\cite{Aaboud:2017nmi}. The RPV gluino has the same signal topology as Eq.~(\ref{eq:topology}), but the cross section for a colour octet is larger than for a colour triplet top partner. In this work we will study various qualitatively different regions of parameter space including:
\begin{itemize}
\item Off-shell $\eta$: $m_\eta \gg m_T$,
\item Bulk on-shell region: $m_\eta \lesssim m_T$,
\item Very light $\eta$: $m_\eta \ll m_T$,
\item Degenerate region: $m_\eta \sim m_T$.
\end{itemize}
We will show that all of these regions are constrained up to a $T$ mass of about 700 to 900 GeV.

We also consider the possibility that the scalar particle ($\eta$) decays to two bottom jets instead of light jets, as might be expected from a Higgs-like scalar. The complete process for this particular decay mode is:
\begin{equation}
pp \rightarrow T\bar{T} \rightarrow jj\eta\eta \rightarrow 2j4b.
\label{eq:B_topology}
\end{equation}
From the presence of the two $b$-jets in the final state, one might expect that the corresponding limit on $T$ would be stronger than in the  light jets case, but in fact the constraints are very similar. Adding $b$-tagging to current search strategies can significantly improve limits on this scenario. 

This paper is structured as follows. In section~\ref{sec:level3} we present the  bounds coming from the latest LHC searches for top partners decaying exclusively to light jets. In particular, we consider searches looking for pairs of resonances decaying to three jets, and searches looking for pairs of di-jet resonances. In section~\ref{subsec:level33} we repeat this exercise for final state containing $b$-jets. In section~\ref{sec:level4} we give a brief summary of the results. This work is augmented by four appendices: in appendix~\ref{app:cms_threeJet_search} we give some details regarding the three-jet CMS resonance search performed at $\sqrt{s} = 13$ TeV, in appendix~\ref{app:ATLAS_diJet_search} we present all the selection requirements for the di-jet ATLAS search applied to our model, in appendix~\ref{app:CMS_threeJet_8TeV_search} we discuss the three-jets CMS search conducted at $\sqrt{s} = 8$ TeV, and in appendix~\ref{app:QCD_section} we give details on how QCD background events are simulated.


\section{\label{sec:level3}Bounds from LHC Searches }
The topologies we consider here consist of resonances that ultimately decay fully hadronically leading to a six-jet final state, not counting initial and final state radiation. Furthermore, we assume that the scalar ($\eta$)  decays promptly; hence, resulting in the absence of any displaced vertices. This multijet final state narrows down the list of possible searches sensitive to this model. The pertinent searches to consider are the ones looking for multiple jets but no missing energy, leptons, or photons. 

\begin{figure}[hbt!]
\centering 
\includegraphics[width=.8\textwidth,origin=c,angle=0]{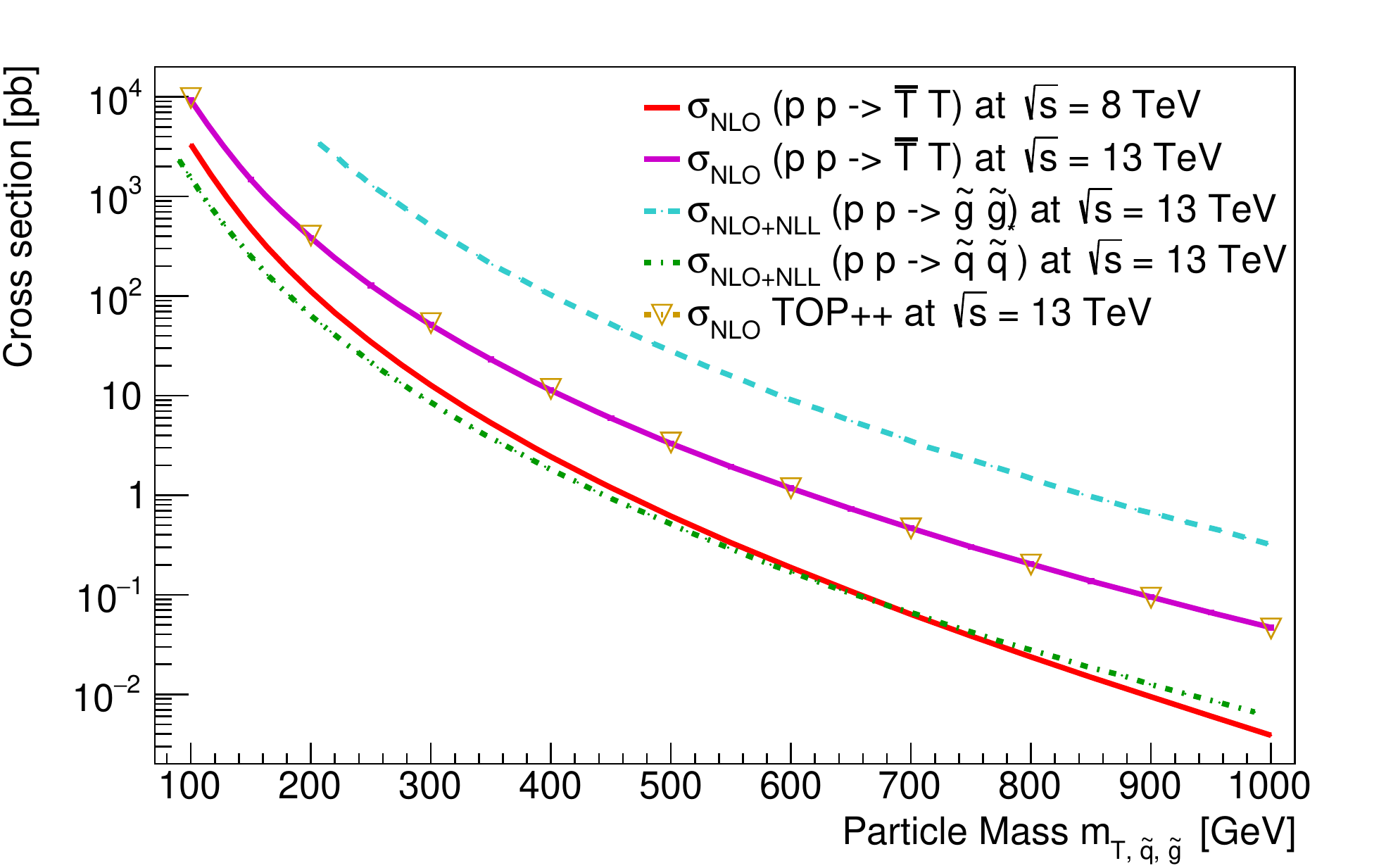}
\caption{\label{fig:myCrossection} Next-to-leading order (NLO) pair production cross section for the top partner as a function of $m_{T}$ at $\sqrt{s} = 8 , 13$ TeV. The cross section was computed using MadGraph5 at next-to-leading order by varying the top quark mass. The figure also contains the NLO top partner pair production cross section at $\sqrt{s} = 13$ TeV computed using TOP++2.0 program for a few benchmark points (triangles). Error bars from varying the renormalization and factorization scales are included but too small to see. The next-to-leading order plus next-to-leading-logarithm (NLL) cross sections for gluino~\cite{Sirunyan:2018duw} and squark~\cite{Aaboud:2017nmi} pair production at $\sqrt{s} = 13$ TeV are also shown for comparison. }
\end{figure}

In order to recast existing searches, we use a few publicly available software packages/tools. The model file for our model was created using the Mathematica package $FeynRules$~\cite{Alloul:2013bka} which was then supplied as an input to MadGraph5 \cite{Alwall:2014hca} for Monte Carlo (MC) event generation. Next, the events were passed to PYTHIA 8 for showering and hadronization~\cite{Sjostrand:2014zea}; subsequently, DELPHES 3~\cite{deFavereau:2013fsa} was used for fast detector simulation and FastJet~\cite{Cacciari:2011ma} was deployed to reconstruct jets.

The top partner pair production cross section was computed at next-to-leading order (NLO) using MadGraph5 by setting the top quark mass to $m_{T}$ and its behaviour as a function of the top partner mass is shown in figure~\ref{fig:myCrossection}. The theoretical cross sections were also computed with the TOP++2.0 program~\cite{Czakon:2011xx} for comparison and were found to be consistent. The benchmark points displayed in figure~\ref{fig:myCrossection} contain statistical and scale uncertainties from renormalization and factorization scale ($\mu_{F}$ and $\mu_{R}$ are varied from 0.5 to 2 times the nominal value). The error bars are too small to observe on the plot. Furthermore, the corresponding values obtained were confirmed by the available literature computations for top partner pair production cross sections~\cite{Aaboud:2018wxv, Aaboud:2018pii, Aaboud:2017qpr, Sirunyan:2018omb}.

\subsection{\label{subsec:level31} CMS Pair-Produced Three-jet Resonances  }
Searches that explore the fully hadronic decay channels with at least six jets (light) jets in the final state have been conducted at the LHC. Older searches include CMS~\cite{Chatrchyan:2012uxa} and ATLAS~\cite{ATLAS:2012dp,Aad:2015lea} searches at $\sqrt{s}=7,8$ TeV that place no constraint on our signal. A similar conclusion was reached by the authors for the double trijet resonances for the composite models in~\cite{Redi:2013eaa} which considers the CMS search~\cite{Chatrchyan:2012uxa}. However, their topology is slightly different since it involves a color octet rather than a singlet $\eta$. The latest multijet search that matches our desired search criteria was conducted by CMS using $35.9 \;\text{fb}^{-1}$ of data collected at a $13$  TeV center-of-mass energy \cite{Sirunyan:2018duw}. The search is designed to look for a pair of particles each decaying to three jets. The analysis interprets the results in the framework of an R-parity violating (RPV) SUSY model where gluinos are pair produced and each decay to three quarks, resulting in a six-jet final state. The search explores a gluino mass range from 200 to 2000 GeV and excludes gluino masses below 1500 GeV at 95$\%$ confidence level. This dedicated analysis focuses on three-jet resonances and takes advantage of Dalitz variables \cite{Dalitz:1954cq} to enhance  signal sensitivity. A distance parameter, sensitive to the symmetry of the jets inside a triplet is defined as:
\begin{equation}
D^{2}_{[3,2]} = \sum_{i > j} \left (  \hat{m}(3,2)_{ij} - \frac{1}{ \sqrt{3} } \right )^{2} \; \;,
\end{equation}
where $\hat{m}(3,2)_{ij}$ is the normalized di-jet invariant masses and is defined as:
\begin{equation}
\hat{m}(3,2)_{ij}^{2} = \frac{ m_{ij}^{2} }{m_{i}^{2} + m_{j}^{2} + m_{k}^{2} + m_{ijk}^{2}} \quad \text{, where }  {i , j , k \in \{1,2,3\}}\;\;.
\end{equation}
Here, $m_{ijk}$ is the triplet invariant mass and $m_{i}$ are the constituent jet masses of the triplet. The complete list of selection criteria used in this particular search are given in table~\ref{table:CMS_selectionTable} in the appendix. To set bounds, the QCD and combinatorial backgrounds are modelled with a monotonically decreasing function, which is optimized in four mass regions. A statistically significant signal-like ``bump'', parametrized by a double Gaussian, is then looked for on top of this background.

\begin{figure}[hbt!]
\centering 
\includegraphics[width=.8\textwidth,origin=c,angle=0]{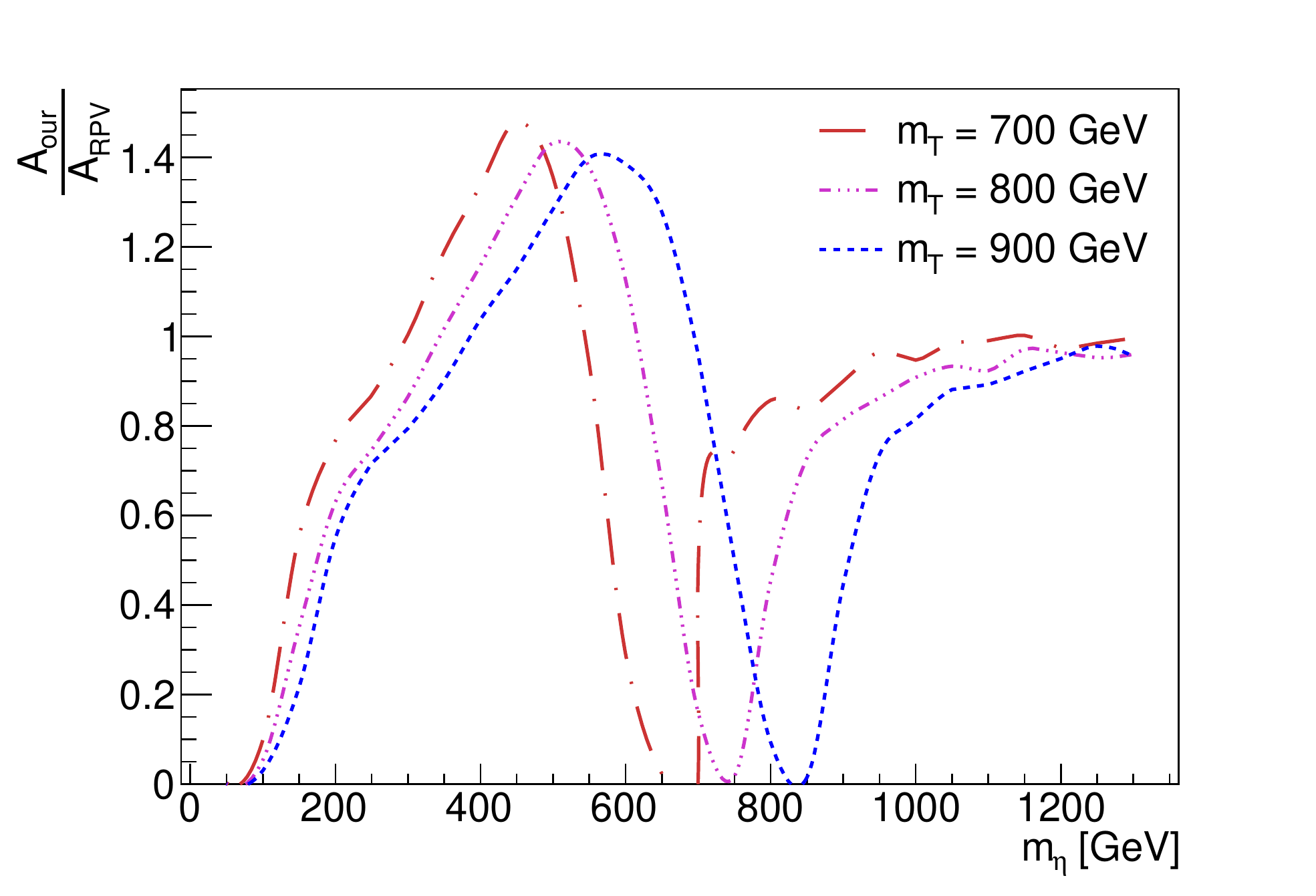}
\caption{\label{fig:CMS_ratioOFaccept} The ratio of acceptance in the recasted CMS search~\cite{Sirunyan:2018duw} for our model over the RPV benchmark model as a function of the scalar mass for a few top partner masses.}
\end{figure}

In order to recast this search and obtain bounds on the parameter space of our model, we first simulated the RPV SUSY topology given in~\cite{Sirunyan:2018duw} with all superpartners except the gluino decoupled. We then simulated our particular model for the corresponding $m_{T}$ ($m_{T} = m_{\tilde{g}}$) for a fixed $m_{\eta}$. We computed the acceptance, defined as the number of correct triplets passing all the selection criteria given in table~\ref{table:CMS_selectionTable} divided by the total number of events generated, for both the original RPV topology and our model. The correct triplets are the ones constructed from the three jets associated with the decay of a gluino (for the CMS topology) or a top partner (for our topology). They have an invariant mass distribution peaked around the resonance mass.\footnote{In our simulation we found the invariant distribution to be slightly skewed towards lower masses, see appendix~\ref{app:cms_threeJet_search} for more details of this discrepancy.}   In order to identify these correct triplets, we require the parton level decay products to be within $\Delta R = 0.3$ from the detector level jet axis. We then rescale the pair production cross section for $\bar{T} T$ shown in figure~\ref{fig:myCrossection} by the ratio of the acceptances as: 
\begin{equation}
\sigma_{rescaled} = \sigma (pp \rightarrow \bar{T}T) \times \frac{ \mathcal{A}_{our} }{\mathcal{A}_{RPV} }\; \;,
\label{eqn:LimitsComputation}
\end{equation} 
where $\mathcal{A}_{our}$ is the acceptance for our topology and $\mathcal{A}_{RPV}$ is the acceptance obtained by simulating the RPV benchmark model in~\cite{Sirunyan:2018duw}.  Because the invariant mass distribution of correct triplets for our topology, including cases where $\eta$ is on-shell, is very similar to the RPV topology, the number of events that our model would produce in an invariant mass peak distinguishable from background is given by $\sigma_{\text{rescaled}} \times \mathcal{A}_{RPV}$.   So a mass point in the $m_{T}-m_{\eta}$ plane is excluded if $\sigma_{rescaled}$ is greater than the observed 95$\%$ upper bound on the cross section $pp \rightarrow \tilde{g} \tilde{g}$ for a given value of $m_{T}$ ($m_{T} = m_{\tilde{g}}$) obtained by the search.

\begin{figure}[hbt!]
\centering 
\begin{subfigure}[b]{0.75\textwidth}
\includegraphics[width=1.0\textwidth,origin=c,angle=0]{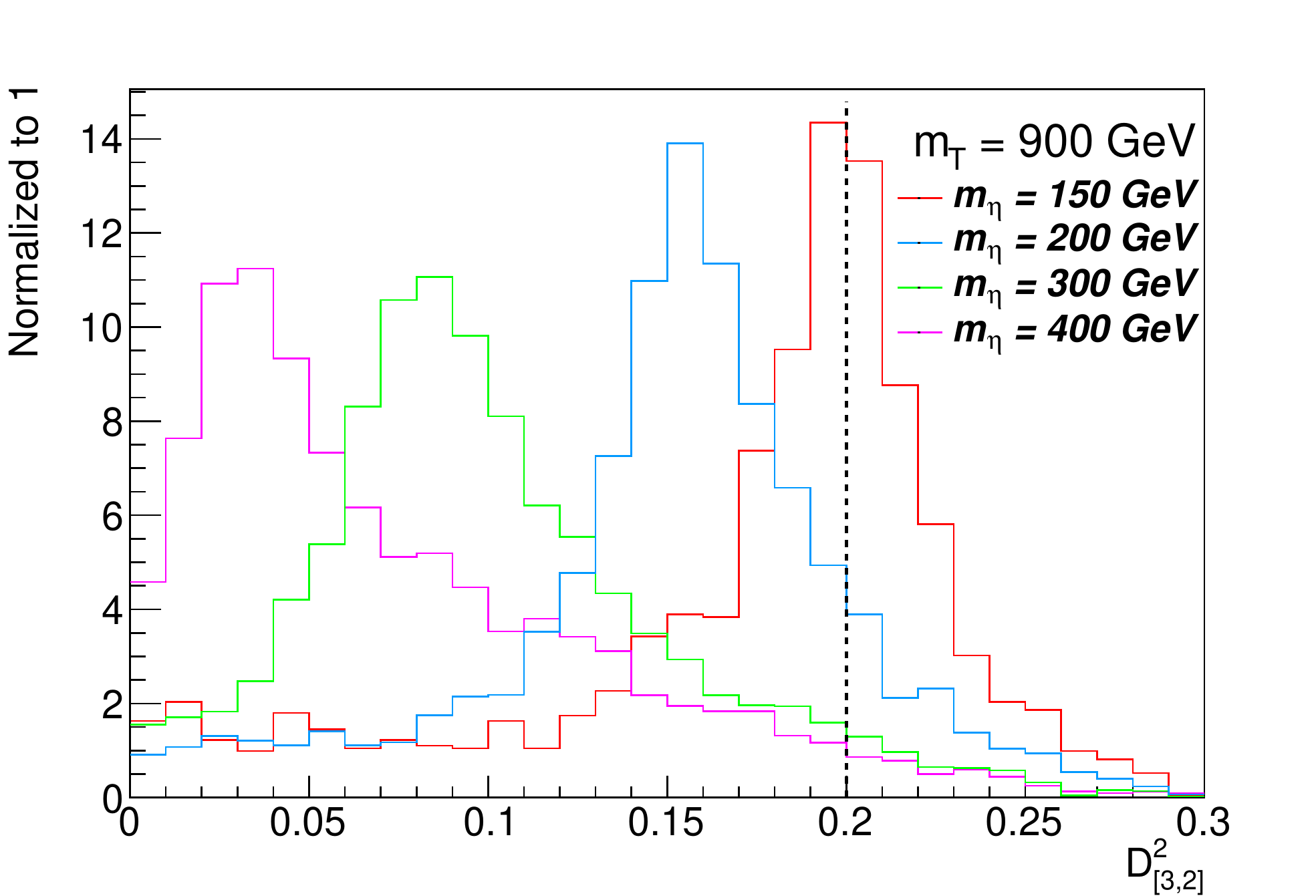}
\caption{  }
\end{subfigure}
\quad
\begin{subfigure}[b]{0.75\textwidth}
\includegraphics[width=1.0\textwidth,origin=c,angle=0]{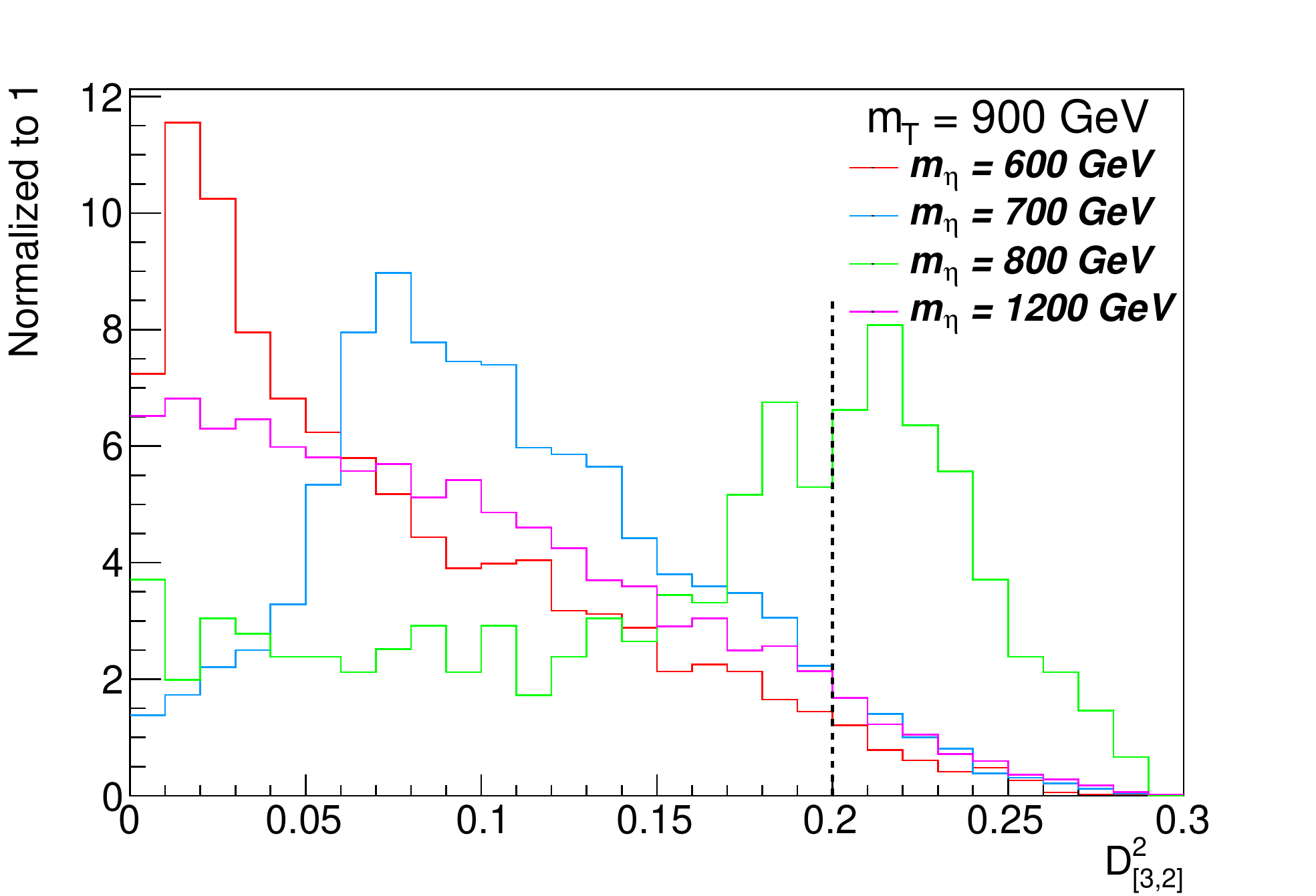}
\caption{  }
\end{subfigure}
\caption{\label{fig:CMS_D32Distribution_mT900} The $D_{[3,2]}^{2}$ variable distributions for signal triplets with top partner mass of 900 GeV and various scalar masses. The black dashed line represent the cut placed on $D_{[3,2]}^{2}$ (accepting events below this value) for this particular top partner mass. }
\end{figure}

To understand the main features of the exclusion regions we obtained, it is instructive to look first at the ratio of acceptance $ \frac{ \mathcal{A}_{our} }{\mathcal{A}_{RPV} }$, which is shown in figure~\ref{fig:CMS_ratioOFaccept} as a function of the scalar mass for various fixed top partner masses. If $m_\eta \gg m_{T}$, then the topology of our model is the same as the RPV gluino decaying to three SM jets and we would expect $\frac{ \mathcal{A}_{our} }{\mathcal{A}_{RPV} } \sim 1$. In this case the bound on $m_T$ can be found by simply comparing $\sigma (pp \rightarrow \bar{T}T)$ with the limit obtained by the CMS Collaboration~\cite{Sirunyan:2018duw}. However, if $m_{\eta} \lesssim m_T$, the scalar $\eta$ is on-shell and the topology is different from the RPV gluino. In particular, the distribution of the $D_{[3,2]}^{2}$ variable changes significantly as $\eta$ goes on-shell and becomes strongly dependent on the mass difference $m_T-m_{\eta}$. This is shown in figure~\ref{fig:CMS_D32Distribution_mT900}  where distributions of the $D_{[3,2]}^{2}$  variable is shown for a fixed $m_{T}= 900$ GeV and several scalar masses.\footnote{Other $m_{T}$ values show similar behaviour.} We find that even with such different distributions, the efficiency of the search remains high in most of the on-shell $\eta$ parameter space because the cut applied on $D_{[3,2]}^{2}$ is relatively high, with the exceptions being for $m_\eta \approx m_T$ and $m_{T}  \gg m_\eta$. The $D_{[3,2]}^{2}$ distributions for the RPV signal are extremely similar to case where $m_\eta \gg m_{T}$ as one would expect which we have alluded to before during the discussion of figure~\ref{fig:CMS_ratioOFaccept}.  

These regions of low efficiencies can be understood as follows: when the mass splitting between $T$ and $\eta$ is small, the jet from the $T\rightarrow j \eta$ decay is soft, resulting in a decrease in $\mathcal{A}_{our}$ from the requirement of six hard jets. This topology however still has a di-jet signature, and can be probed by the di-jet search discussed in the next section. If $ m_{T}  \gg m_\eta$, the scalars are produced with very high boost and the jets resulting from their decay $\eta \rightarrow j j$ will often merge into a single jet leading once again to reduced sensitivity, but also having a di-jet like topology.

\begin{figure}[hbt!]
\centering 
\includegraphics[width=.95\textwidth,origin=c,angle=0]{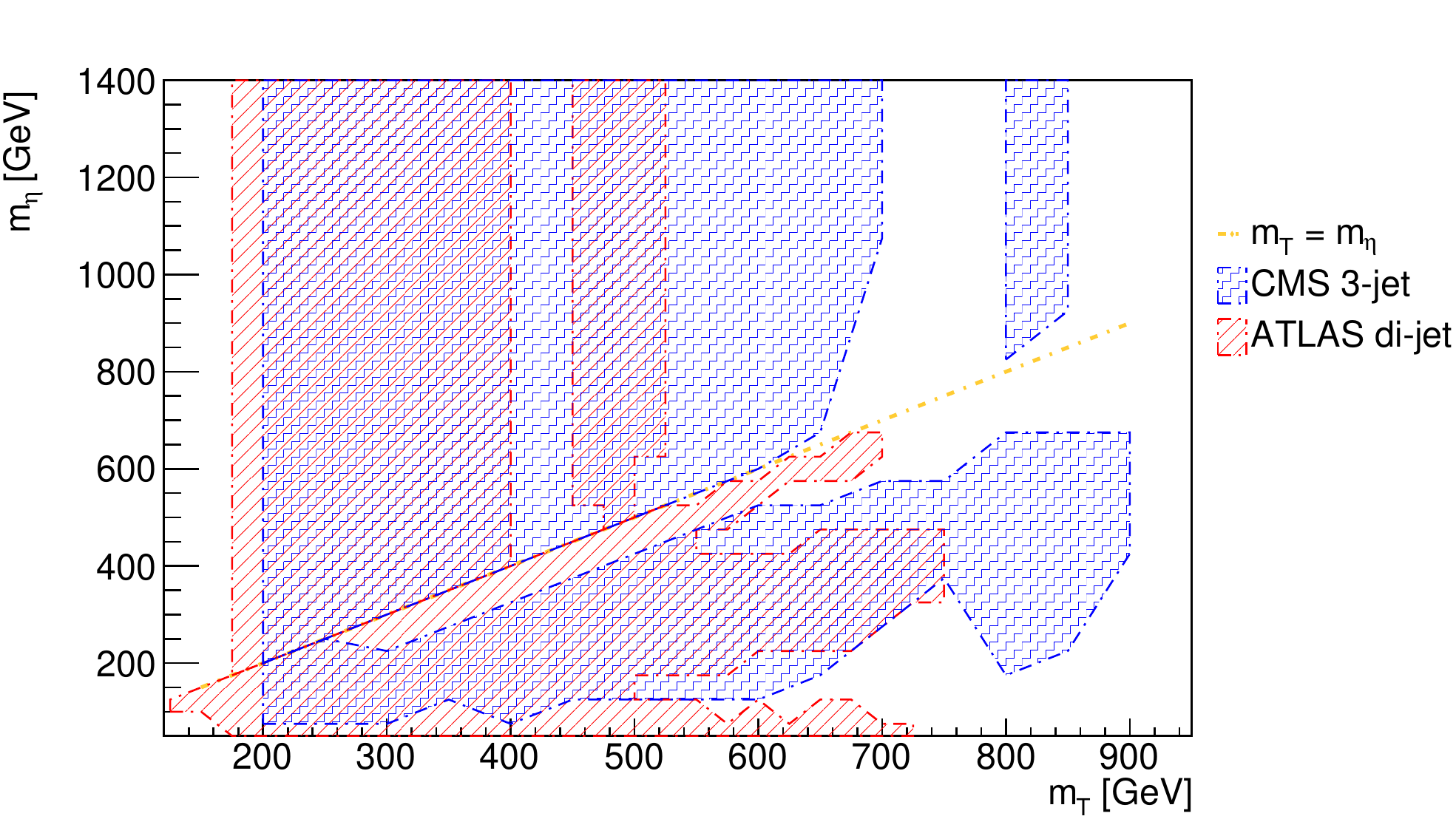}
\caption{\label{fig:Exclusion_Final}The CMS and ATLAS bounds on the fully hadronic decay mode (to light jets) of the top partner. The shaded regions are excluded by the current searches conducted at the LHC \cite{Aaboud:2017nmi, Sirunyan:2018duw}.}
\end{figure}

Performing a grid search for all mass points in the  $m_{T}-m_{\eta}$ plane with increments of 25 GeV, we obtain the exclusion regions shown by the blue shaded area of figure~\ref{fig:Exclusion_Final}. We see that the dips in figure~\ref{fig:CMS_ratioOFaccept} translate to holes in the sensitivity of the CMS search for $m_T \sim m_\eta$ and for $m_\eta \ll m_T$. We also see that the bounds in the bulk on-shell region $m_T \gtrsim m_\eta$ are stronger than the off-shell region $m_T \lesssim m_\eta$. The isolated exclusion near $m_{T} \sim 800$ GeV in the off-shell region can be attributed to a downward fluctuation of the background in that region.


\subsection{\label{subsec:level32} ATLAS Pair-Produced Di-Jet Resonances  }
As  mentioned in the previous section, searches that look for pair-produced di-jet resonances can constrain this scenario in the region where $\eta$ is on-shell.\footnote{We will show later that this search also has significant sensitivity in the off-shell region.}  It is especially useful  when $T$ and $\eta$ are close in mass or the scalar is very light as the three-jet search is not sensitive in those regions. Such a search was conducted by ATLAS at $\sqrt{s} = 13$ TeV with an integrated luminosity of 36.7 $\text{fb}^{-1}$~\cite{Aaboud:2017nmi}. It explores coloured resonances that are pair-produced and that each decay to two jets, giving rise to a four-jet final state. The results of the analysis are interpreted in a simplified R-parity violating SUSY model where the top squark is the lightest supersymmetric particle and decays promptly into two quarks ($\tilde{t} \rightarrow \bar{q}_{j} \bar{q}_{k} $). The search explores the region $100 \text{ GeV } < m_{\tilde{t} } < 800 \text{ GeV}$ and excludes top squark masses in the range $100 \text{ GeV} < m_{\tilde{t} } < 410 \text{ GeV}$ at 95$\%$ CL. The list of selection criteria is given in table \ref{table:ATLAS_selectionTable} in the appendix. Since it is expected that the resonances are produced with high transverse momentum, their decay products will be located close to each other. As such, the four leading jets are paired using an angular distance:

\begin{equation}
\Delta R_{min} = \text{min } \left \{ \sum_{i=1}^{2} \abs{\Delta R_{i}  - 1 }  \right \} \; \;,
\end{equation}
where $\Delta R_{i} = \sqrt{  \Delta \phi_{i}^{2} + \Delta \eta_{i}^{2}  }  $ is the distance between the two jets in $i^{th}$ pair. The two jet pairs selected must minimize $\Delta R_{min}$ and satisfy the $\Delta R_{min}$ cut. Signal jets are expected to be produced in the central region so putting a cut on $\abs{ \cos{\theta^{\ast}} }$, where $\theta^*$ is the angle that either of the resonances makes with the beamline in the center-of-mass frame, is beneficial. Finally, the masses of the resonances are expected to be equal; hence, their invariant mass differences would be an ideal discriminant between signal and background. As such, the mass asymmetry ($\mathcal{A}_{m}$), defined as:
\begin{equation}
\mathcal{A}_{m}  = \frac{ \abs{ m_{1} - m_{2} } }{ m_{1} + m_{2} } \; \;,
\label{eq:Am}
\end{equation}
where $m_{1}$ and $m_{2}$ are the invariant masses of the two reconstructed di-jet pairs is required to be small. To set bounds, the ATLAS collaboration employed a modified frequentist approach using the CL$_{s}$~\cite{Junk:1999kv, Read_2002, Cowan:2010js} technique and a profile likelihood ratio as the test statistics. For each mass hypothesis, a counting experiment is performed in a window around the average mass of the two reconstructed resonances. The dominant source of background comes from QCD multijet production and is estimated directly from the data with a method that predicts both the normalization and the shape of the average di-jet mass distribution~\cite{Aaboud:2017nmi}.

\begin{figure}[hbt!]
\centering 
\begin{subfigure}[b]{0.47\textwidth}
\includegraphics[width=1.1\textwidth,origin=c,angle=0]{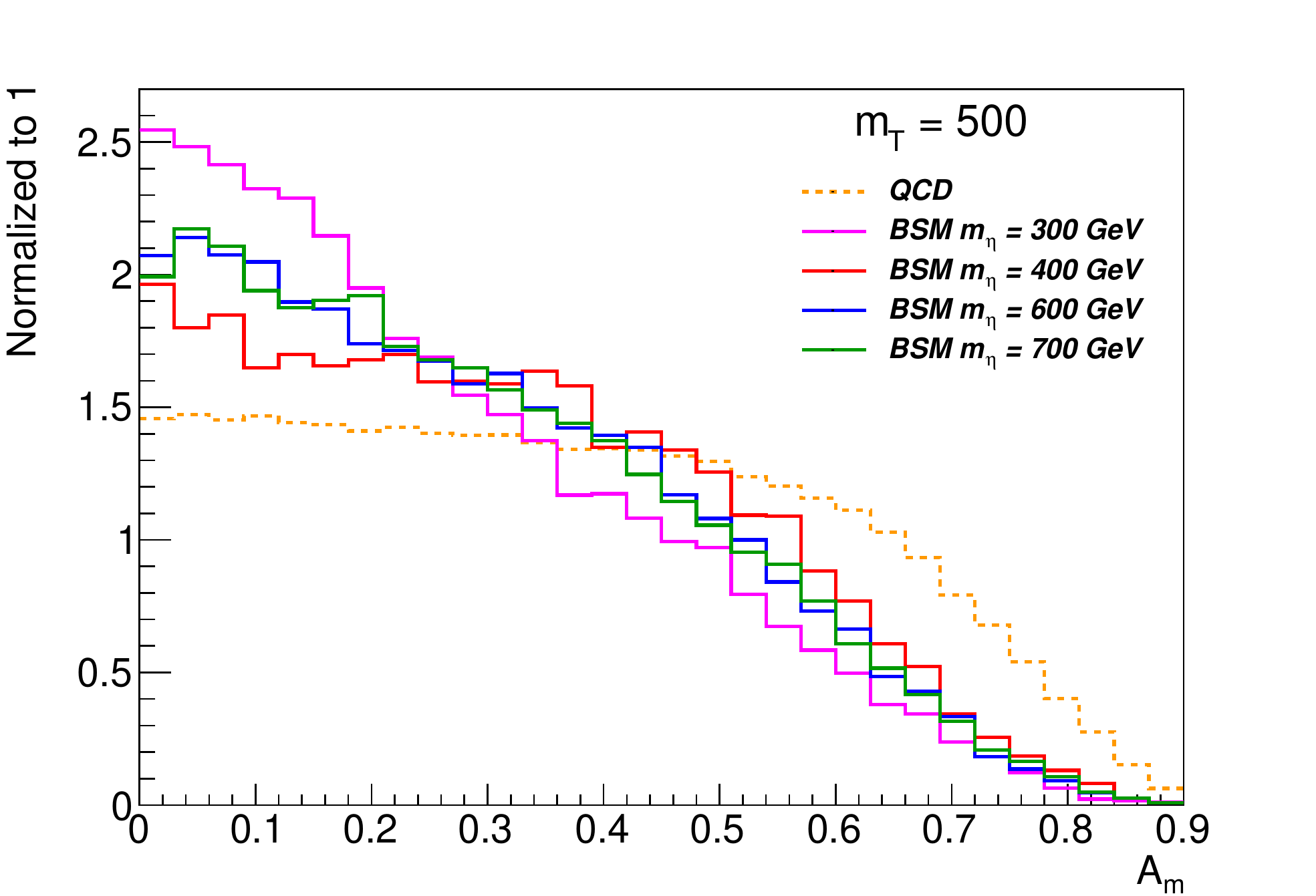}
\caption{ Passing only $p_{T}$ cut }
\end{subfigure}
\quad
\begin{subfigure}[b]{0.47\textwidth}
\includegraphics[width=1.1\textwidth,origin=c,angle=0]{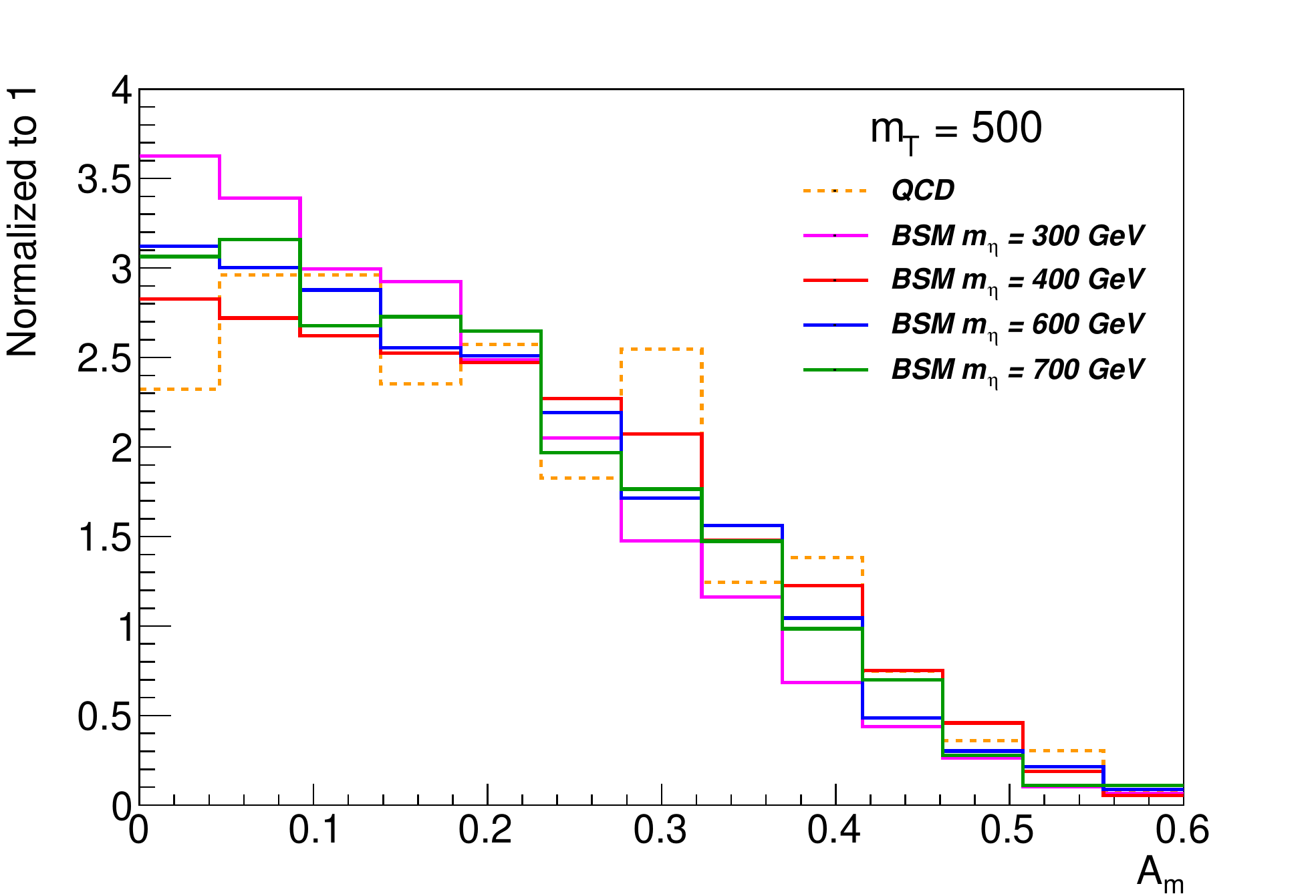}
\caption{ Passing  $p_{T}$ and $\Delta R_{min}$ cuts }
\end{subfigure}
\\
\begin{subfigure}[b]{0.5\textwidth}
\includegraphics[width=1.15\textwidth,origin=c,angle=0]{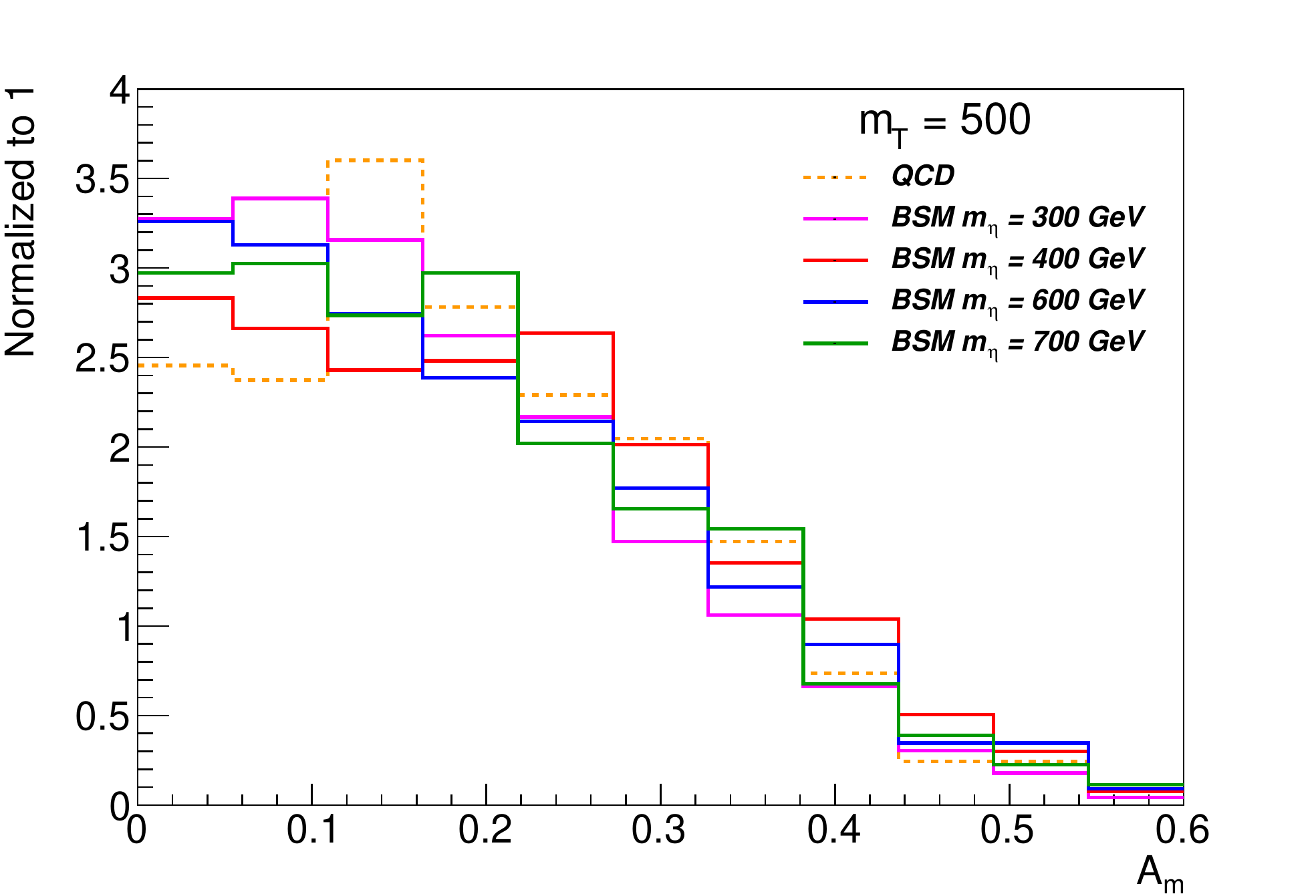}
\caption{ Passing all but $\mathcal{A}_{m}$ cuts }
\end{subfigure}
\quad

\caption{\label{fig:Am_dijet_Final_Distribution} The $\mathcal{A}_{m}$ spectrum passing the various selection criteria as highlighted in the ATLAS search~\cite{Aaboud:2017nmi} for $m_{T}=500$ GeV. The distributions of the (a) are for events passing only the $p_{T}$ requirement, the (b) satisfies both $p_{T}$ and $\Delta R_{min}$ requirements, and the (c) have all except $\mathcal{A}_{m}$ cuts applied. }
\end{figure}

In order to recast the search, we simulated our particular model for the corresponding $m_{T}$ ($m_{T} = m_{\tilde{t}}$)  for a fixed $m_{\eta}$. We computed the efficiency of the search for our model ($\epsilon_{our}$), defined as the number of events satisfying all the cuts in table~\ref{table:ATLAS_selectionTable} plus an invariant mass window cut over the total number of simulated events. The number of signal events is then given as follows:
\begin{equation}
 N_{s} = \mathcal{L}_{luminosity} \times \sigma (pp \rightarrow \bar{T}T) \times \epsilon_{our} \; \;.
\end{equation}
We computed this number for various window mass cuts, corresponding to different stop masses considered in~\cite{Aaboud:2017nmi}. Taking the number of background and observed events and their uncertainties in the different windows from tables 3 and 5 of~\cite{Aaboud:2017nmi}, we computed the confidence level:
\begin{equation}
CL_{s} = \frac{ P_{s+b} ( X \leq X_{obs} )  }{ P_{b} ( X \leq X_{obs} ) } = \frac{CL_{s+b}}{CL_{b}} \; \; .
\end{equation}
Here $CL_{s+b}$ is the confidence level for excluding the possibility of simultaneous presence of signal and background while $CL_{b}$ is the probability that the test statistic is less than or equal to that observed in the data (assuming only the presence of background)~\cite{Read_2002,Cowan:2010js,Junk:1999kv}. Then a mass point is excluded at 95$\%$ confidence level if $(1-CL_{s})\times 100 \%$ is greater than 95$\%$. The parameter space excluded by the ATLAS search is given by the red shaded region of figure  \ref{fig:Exclusion_Final}. It extends up to top partner masses of $\sim$ 750 GeV.

The di-jet search allows us to close some of the gaps that remain in the three-jet resonance search, namely the regions where $m_\eta \approx m_T$ or  $m_\eta \ll m_T$.  The gap in di-jet exclusion curve for $400 \text{ GeV} \leq m_{T} \leq 425 \text{ GeV} $ and $m_{T} < m_{\eta}$ can be explained by a downward fluctuation in the background in that region as illustrated in figure 9 (a) of the ATLAS search~\cite{Aaboud:2017nmi}. 

While the ATLAS search is primarily designed for di-jet topologies we also find reasonably good efficiency in the off-shell region where there is in fact no di-jet resonance, placing an exclusion up to $m_{T} \leq 525$ GeV in that region. 
In particular, one would expect the $A_m$ cut in Eq.~\ref{eq:Am} would be very inefficient when there is no 2-jet resonance. However, we find that the $\Delta R_{min}$ and $\abs{ \cos (\theta^{\ast})}$ selection criteria sculpt the $A_{m}$ distribution which becomes similar for the on-shell and off-shell cases. This is shown in figure~\ref{fig:Am_dijet_Final_Distribution} where $A_m$ is plotted after various cuts.

For very light top partner masses $m_{T} \lesssim 200$ GeV, searches at previous experiments may be sensitive. For example, a three-jet hadronic resonance search was performed at $\sqrt{s}=1.96$ GeV at the Collider Detector at Fermilab (CDF)~\cite{Aaltonen:2011sg} and excluded gluinos below 144 GeV. Multijet searches performed at LEP has excluded the neutralino decaying to three jets for masses up to about 100 GeV~\cite{Barate:1999fc, Heister:2002jc}. There does remain a gap for $m_T \simeq m_t$ near the top quark. It may be possible to exclude this region of parameter space using the measurement of the all hadronic top quark decay~\cite{Chatrchyan:2013ual,Aad:2014zea,Khachatryan:2015fwh,Aaboud:2017mae,ATLAS:2020aub}, but a detailed analysis is beyond the scope of this work.  

\section{\label{subsec:level33} Heavy Flavour Scenario  }
\begin{figure}[h]
\centering 
\includegraphics[width=14cm, height=8cm]{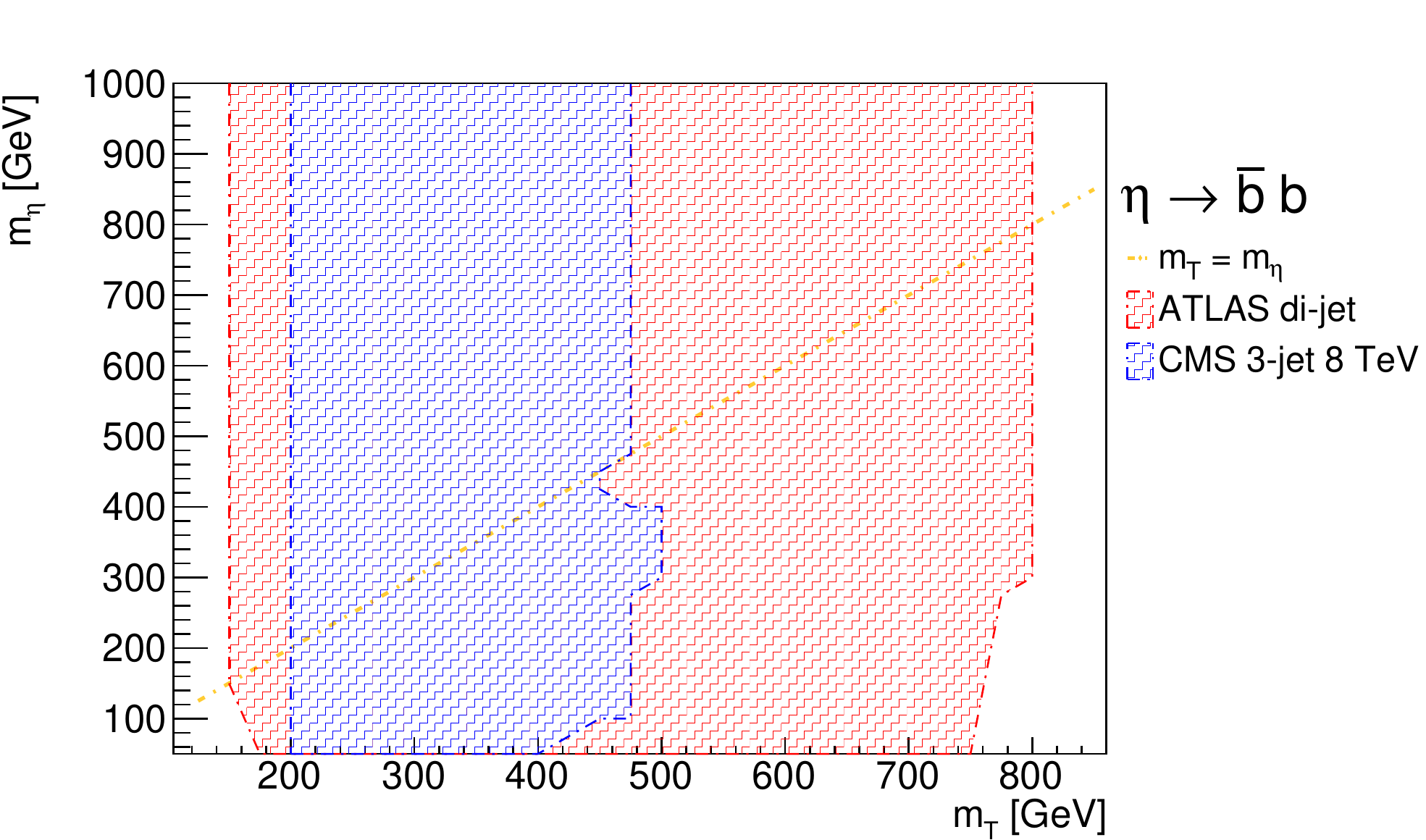}
\caption{\label{fig:Exclusion_bTag_Final} The CMS and ATLAS bounds on the fully hadronic decay mode (to heavy flavour jets) of the top partner. The shaded regions are excluded by the current searches conducted at the LHC \cite{Chatrchyan:2013gia, Aaboud:2017nmi}.}
\end{figure}

In this section we study the case where the scalar $\eta$ decays to two $b$-jets: $pp \rightarrow T\bar{T} \rightarrow jj\eta\eta \rightarrow 2j 4b$. This decay topology is what one would expect if, for example,  the $\eta$ coupled to fermions proportionally to their masses and was lighter than twice the top mass. 
 In principle such a final state allows for better discrimination from QCD multijet events. However, we find that existing searches do not give much stronger constraints. We considered the fully hadronic ATLAS R-parity-violating multijet searches~\cite{Aad:2015lea} and~\cite{Aaboud:2018lpl} performed at $\sqrt{s} = 8$ TeV and 13 TeV respectively, the 13 TeV ATLAS di-jet search~\cite{Aaboud:2017nmi} and the heavy flavour three-jet resonance CMS search~\cite{Chatrchyan:2013gia} conducted at  $\sqrt{s} = 8$ TeV. All of these searches have a $b$-tagging requirement in some of their signal regions, but only the CMS three-jet resonance and the ATLAS di-jet searches were found to place a limit on this scenario. The selection criteria for the heavy flavour three-jet resonance CMS search\footnote{The search also requires at least 6 jets and at least one $b$-jet.} are presented in table~\ref{table:CMS_8TeV_selectionTable} (in the appendix) and the limits on the parameter space are computed using the same method we used to recast~\cite{Sirunyan:2018duw}  (see equation~\ref{eqn:LimitsComputation}). The region of parameter space excluded at 95$\%$ confidence level by this particular CMS search is presented by the blue shaded region of figure~\ref{fig:Exclusion_bTag_Final}. 

We also recasted the ATLAS di-jet search~\cite{Aaboud:2017nmi}, which requires at least two $b$-tagged jets in addition to the window cut around the average invariant mass of the di-jets with additional selection criteria given in table~\ref{table:ATLAS_selectionTable} (in the appendix). Computing the exclusion limits using the CL$_{s}$ method~\cite{Read_2002, Cowan:2010js, Junk:1999kv}, we obtain the red shaded region in figure~\ref{fig:Exclusion_bTag_Final}. The presence of at least two $b$-jet in the reconstructed resonances reduces the combinatoric background coming from the two jet pairing. Hence, the search often selects the di-jet pairs corresponding to the scalar resonance, making the search more effective in both on-shell and off-shell regions in comparison to the light jet scenario.

The searches considered in section~\ref{sec:level3} do not apply $b$-tags, but they also do not veto events with $b$-jets, so the exclusion regions shown in figure~\ref{fig:Exclusion_Final} also apply to the $b$-rich scenario considered in this section. Therefore, comparing figures~\ref{fig:Exclusion_Final} and~\ref{fig:Exclusion_bTag_Final}, we see that adding $b$-tagging only provides new exclusions for $m_T$ between about 700 and 800 GeV and some values of $m_\eta$. The ATLAS search~\cite{Aaboud:2017nmi}, however, only investigates $m_{\tilde{t}}$ up to 800 GeV which is almost entirely excluded.\footnote{With the exception of the very light scalars for $m_{T} \leq 150$ GeV and low scalar masses for $m_{T} \geq$ 775 GeV.} So in principle, the ATLAS di-jet search could have excluded larger $m_{T}$. Due to the steeply decreasing cross section, shown in figure~\ref{fig:myCrossection}, it is unlikely that extending the search would have resulted in an exclusion region significantly above 800 GeV. 

The $\sqrt{s}=13$ TeV three light jets resonance search from CMS analyzed in section~\ref{subsec:level31} places the strongest constraints on three-jet resonances. If this search included a signal region with $b$-tagging, it could also improve constraints on the $b$-rich scenario. Here we give a rough appraisal of the potential improvement. 
First, we estimate the total number of QCD multijet events by simulating $pp \longrightarrow jjjj$ using MadGraph5 interfaced with Pythia8 and Delphes (more detail on the simulation is given in appendix~\ref{app:QCD_section}). The acceptance ($\mathcal{A}_{QCD}$) is obtained by applying all the selection criteria for the three-jet CMS search and we approximate the number of background events to be: $b= \mathcal{L}_{luminosity} \times \sigma (pp \rightarrow jjjj )\times \mathcal{A}_{QCD}$. We set the minimum parton level $p_T$ for the simulated QCD events to be 100 GeV in order to have enough events for our rough approximation. In a similar way we compute $s$, the number of signal events using simulation of our model. We then compare the $s/\sqrt{b}$ values obtained without $b$-tagging to the case where with 2 $b$-tags included in the cuts, and find that $s/\sqrt{b}$ is roughly a factor of three larger. The three-jet CMS search  excludes $m_T \lesssim 900$ GeV for the light jet case, so our rough estimate is that including a $b$-tag requirement could exclude the $b$-jet topology up to $m_T \sim 1050$ GeV as the ratio of top partner cross section for $900$ GeV and $1050$ top partner masses is also $\sim 3$. Requiring at least 4 $b$-tags could further increase the expected exclusion to $\sim 1300$ GeV. We summarize the improvements for different numbers of $b$-tags in table~\ref{table:CMS_massSensitivity_table}. 

In a similar manner, we can roughly estimate the effect of extending the ATLAS di-jet search to higher masses  by computing $R  \equiv \sigma_{rescaled } / \sigma_{obs}  $ at the cut off region ($m_{T} = 800$ GeV)  where $\sigma_{obs}$ is the observed $95 \%$ CL from ATLAS di-jet search for a given mass and $\sigma_{rescaled }$ is the rescaled cross section as defined in equation~\ref{eqn:LimitsComputation}. The corresponding value for $R$ is approximately 3 which points to a limit of $\sim m_T = 950$ as the ratio of cross section for top partner masses of $950$ and $800$ GeV is $\sim 3$.

\begin{table}[h]
\caption{The approximate improvement in  $s/\sqrt{b}$ and mass reach as a function of increasing number of $b$-tags, $N_{b}$. The first line is the actual limit, while the subsequent lines are estimated potential improvements. The $s/\sqrt{b}$ values are computed at the cut off region of $m_{T} = 900$ GeV for the three-jet CMS search~\cite{Sirunyan:2018duw} with 35.9 fb$^{-1}$.  }
\centering
\resizebox{\columnwidth}{!}{
\begin{tabular}{c |c c c }
\hline\hline
 B-tagging requirement &  $\frac{s}{\sqrt{b}}$ & Improvement & Mass Sensitivity [GeV] \\ [0.5ex] 
\hline
$N_{b} \geq 0$  & $0.31$ &  - &  $900$  \\
\hline
$N_{b} \geq 1$  & $0.58$ &  $1.86$ &  $1000$  \\
$N_{b} \geq 2$  & $1.04$ &  $3.33$ &  $1050$   \\
$N_{b} \geq 3$  & $2.30$ &  $8.36$ &  $1200$  \\
$N_{b} \geq 4$  & $4.66$ &  $14.89$ &  $1300$ \\
\hline
\end{tabular}
}
\label{table:CMS_massSensitivity_table}
\end{table}


\section{\label{sec:level4}Conclusion }
Although there has been a significant experimental search program for top partner pair production with $t h, b W$ and $tZ$ decay modes, to our knowledge there has not been any studies to explore the all light jet decay mode. In this work we have recasted the latest available LHC searches that can impose significant constraints on the parameter space of models where the top partner decays to light jets.  Our results are shown in figures~\ref{fig:Exclusion_Final} and~\ref{fig:Exclusion_bTag_Final} for models with final states containing only light jets and for final states containing 4 $b$-jets respectively. Top partner masses are generally excluded up to $m_T \sim 900$ GeV, but there are a few gaps in the $m_T-m_{\eta}$ plane for lighter $m_T$. Because the three-jet resonance search we recast focused on a resonance that decays through an off-shell scalar, it might be possible to obtain better limit in the on-shell $\eta$ region by designing a search specifically for that topology. Furthermore we found that existing searches do not provide significantly better constraints for the case where the final states contain $b$-jet, but those limits could be improved by adding $b$-tagging requirements to the 13 TeV three-jet resonance search.

\acknowledgments
We thank Jack Collins and Jesse Thaler for helpful discussions. This work was supported in part by the Natural Sciences and Engineering Research Council of Canada (NSERC). 

\appendix
\section{CMS Three Jet Resonance Search}
\label{app:cms_threeJet_search}
The 13 TeV CMS search \cite{Sirunyan:2018duw} requires an event to contain at least six jets with $\abs{\eta} < 2.4$. The jet reconstruction is performed using the anti-$k_{t}$ algorithm~\cite{Cacciari:2008gp} with a radius parameter of $R=0.4$. The list of all the selection criteria used in this particular search is given in table \ref{table:CMS_selectionTable}. The analysis employs the jet-ensemble technique \cite{Aaltonen:2011sg, Chatrchyan:2012uxa}, which takes the six highest $p_{T}$ jets in a given event and group them into 20 unique triplets. For signal, at most 2 of these triplets per event corresponds to the pair produced gluino decay while the rest contributes to combinatoric background which are referred to as ``incorrect'' triplets. Consequently the acceptance is defined as the ratio of the correct triplet over the total number of triplets (20) in the event. Furthermore, an event-level variable $D_{[(6,3)+(3,2)]}^{2}$ is defined in order to characterize the angular spread of the six constituent jets insides a pair of triplets. The six-jet distance measure is defined as:
\begin{equation}
D_{[(6,3)+(3,2)]}^{2} = \sum_{i < j < k} \left (  \sqrt{ \hat{m}(6,3)_{ijk}^{2} + D^{2}_{[3,2], i j k} }   - \frac{1}{ \sqrt{20} } \right )^{2}  \; \;,  
\end{equation}
where $\hat{m}(6,3)_{ijk}^{2} = \frac{ m_{ijk}^{2} }{ 4\cdot m_{ijklmn}^{2} + 6 \sum_{i} m_{i}^{2}  }$ with ${i,j,k,l,m,n \in \{ 1,2,...,6\}}$ and $m_{ijklmn}$ the invariant mass of the six highest $p_{T}$ jets. For a new particle decaying to three-jets, the jets produced would be uniformly distributed in a detector resulting in $\hat{m}(6,3)_{ijk}^{2}$ approximately $1/20$. While the jets from the QCD are usually grouped together giving $\hat{m}(6,3)_{ijk}^{2}$ close to zero or one. 

\begin{table}[h]
\caption{The list of selection criteria with the direction of the cuts and the mass ranges analyzed by the CMS search~\cite{Sirunyan:2018duw}.}
\centering
\resizebox{\columnwidth}{!}{
\begin{tabular}{c c c c c c c c }
\hline\hline
Gluino mass range [GeV]  & Jet $p_{T}$ [GeV] &  $H_{T}$ [GeV]  &   Sixth Jet $p_{T}$ [GeV]  & $D^{2}_{ [(6,3) + (3,2) ] } $  & $A_{m} $ & $\Delta$ [GeV]  & $D^{2}_{[3,2]}$ \\ [0.5ex] 
\hline
200-400 & $>30$ & $>650$ & $>40$ & $<1.25$ &  $<0.25$ & $>250$ & $<0.05$ \\
400-700 & $>30$ & $>650$ & $>50$ & $<1.00$ &  $<0.175$ & $>180$ & $<0.175$ \\
700-1200 & $>50$ & $>900$ & $>125$ & $<0.9$ &  $<0.15$ & $>20$ & $<0.2$ \\
1200-2000 & $>50$ & $>900$ & $>175$ & $<0.75$ &  $<0.15$ & $> -120$ & $<0.25$ \\
\hline
\end{tabular}
}
\label{table:CMS_selectionTable}
\end{table}

\begin{figure}[b!]
\centering 
\begin{subfigure}[b]{0.45\textwidth}
\includegraphics[width=1.1\textwidth,origin=c,angle=0]{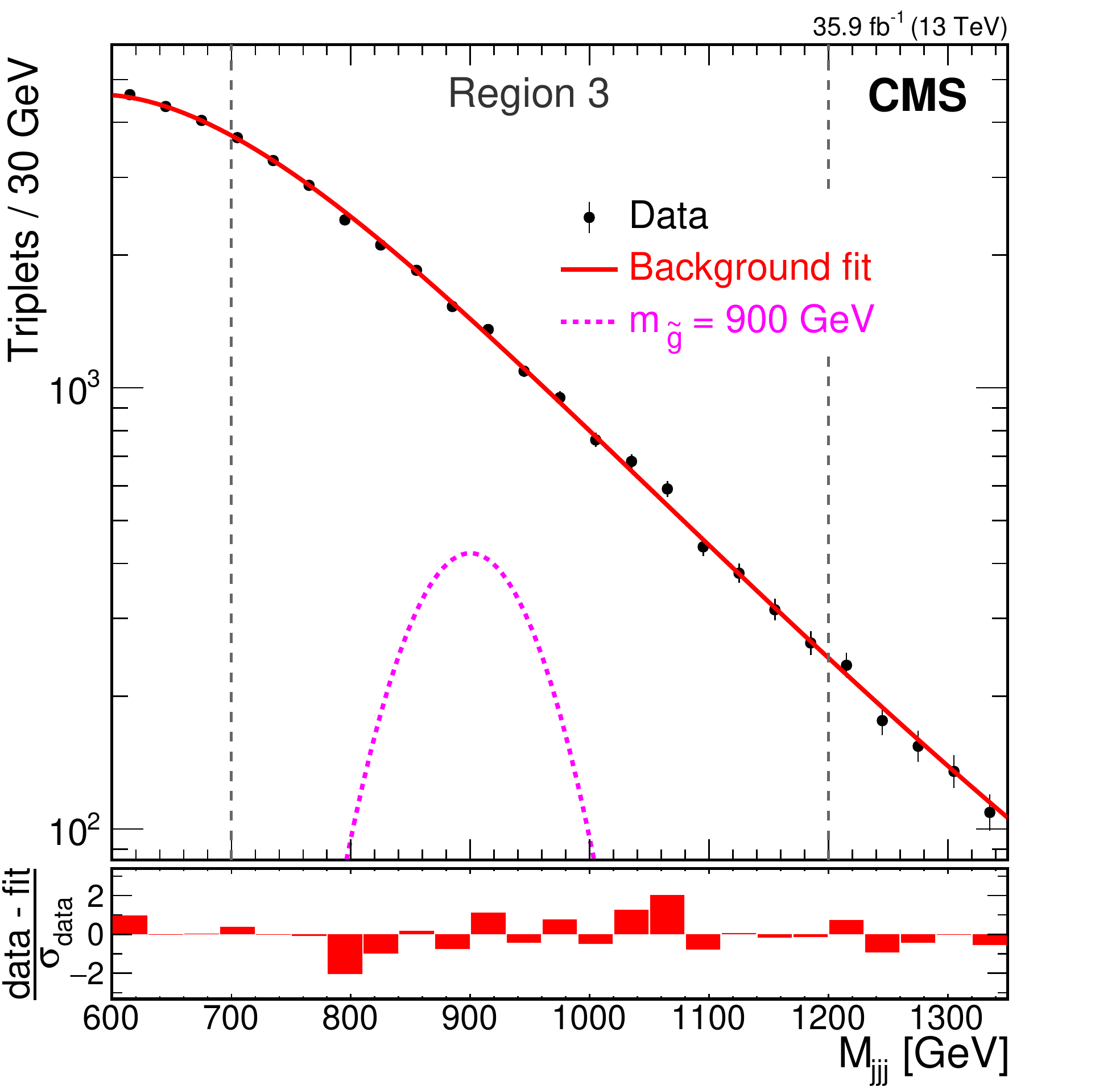}
\caption{  }
\end{subfigure}
\quad
\begin{subfigure}[b]{0.45\textwidth}
\includegraphics[width=1.1\textwidth,origin=c,angle=0]{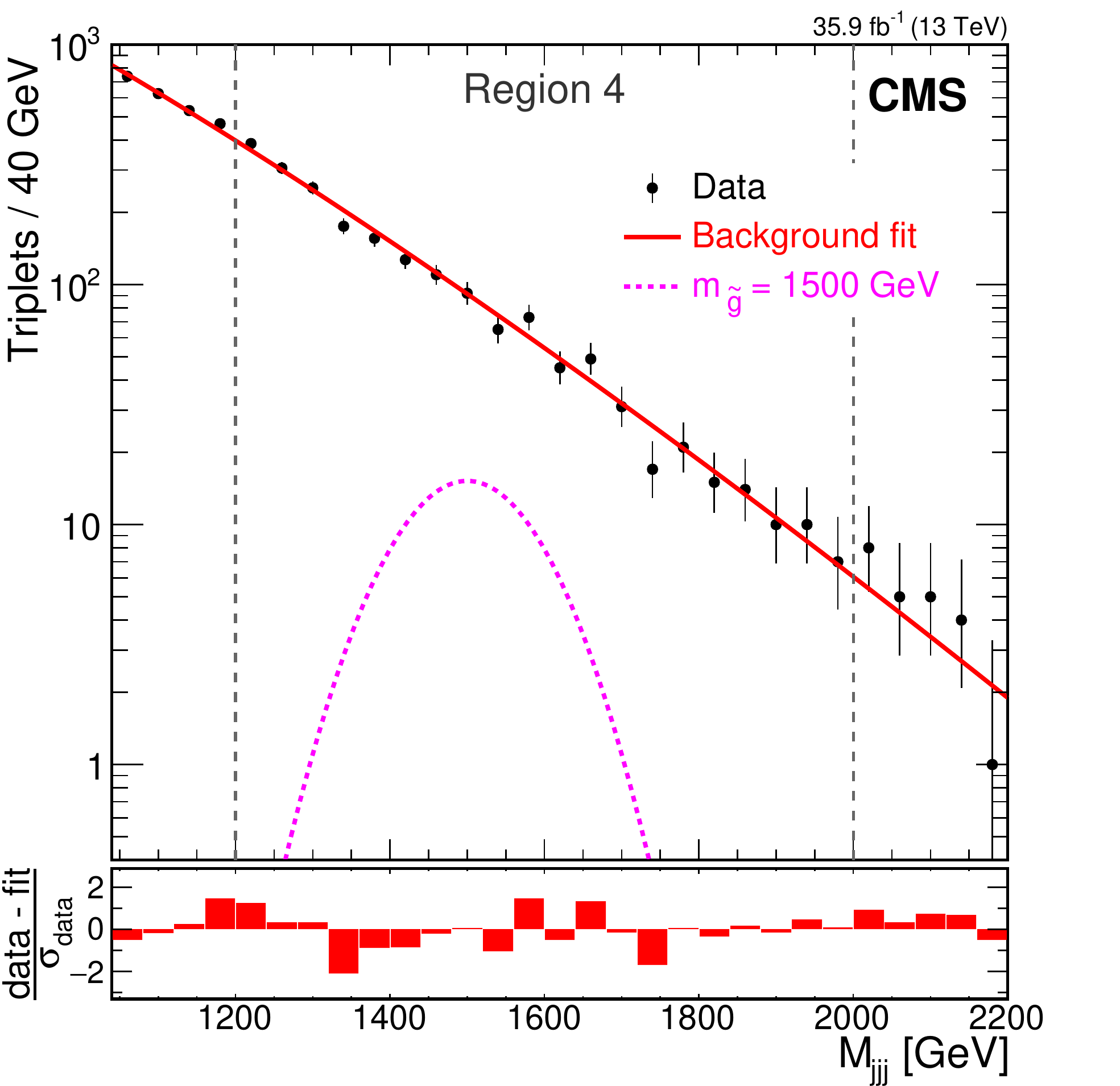}
\caption{  }
\end{subfigure}
\\
\begin{subfigure}[b]{0.45\textwidth}
\includegraphics[width=1.1\textwidth,origin=c,angle=0]{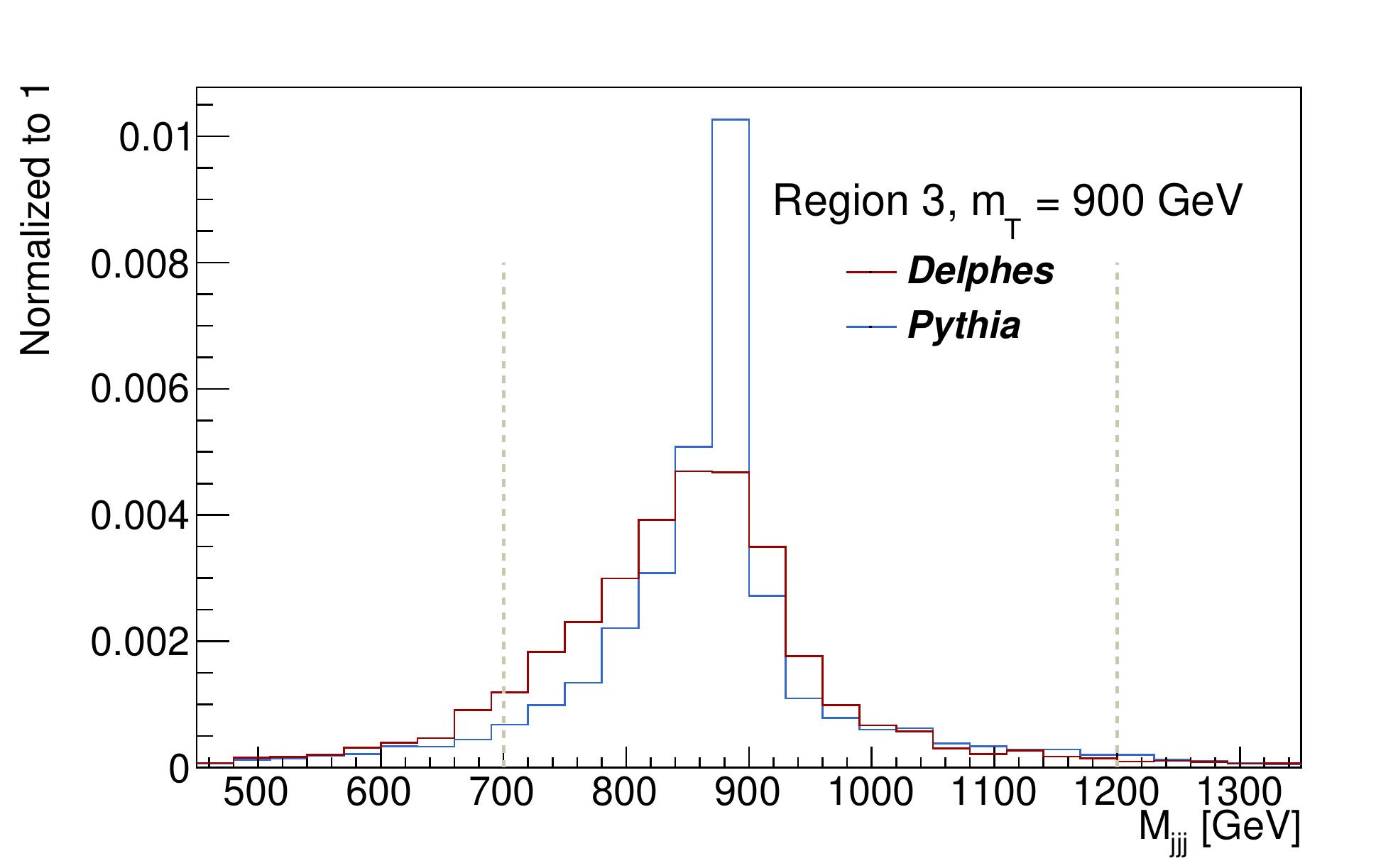}
\caption{  }
\end{subfigure}
\quad
\begin{subfigure}[b]{0.45\textwidth}
\includegraphics[width=1.1\textwidth,origin=c,angle=0]{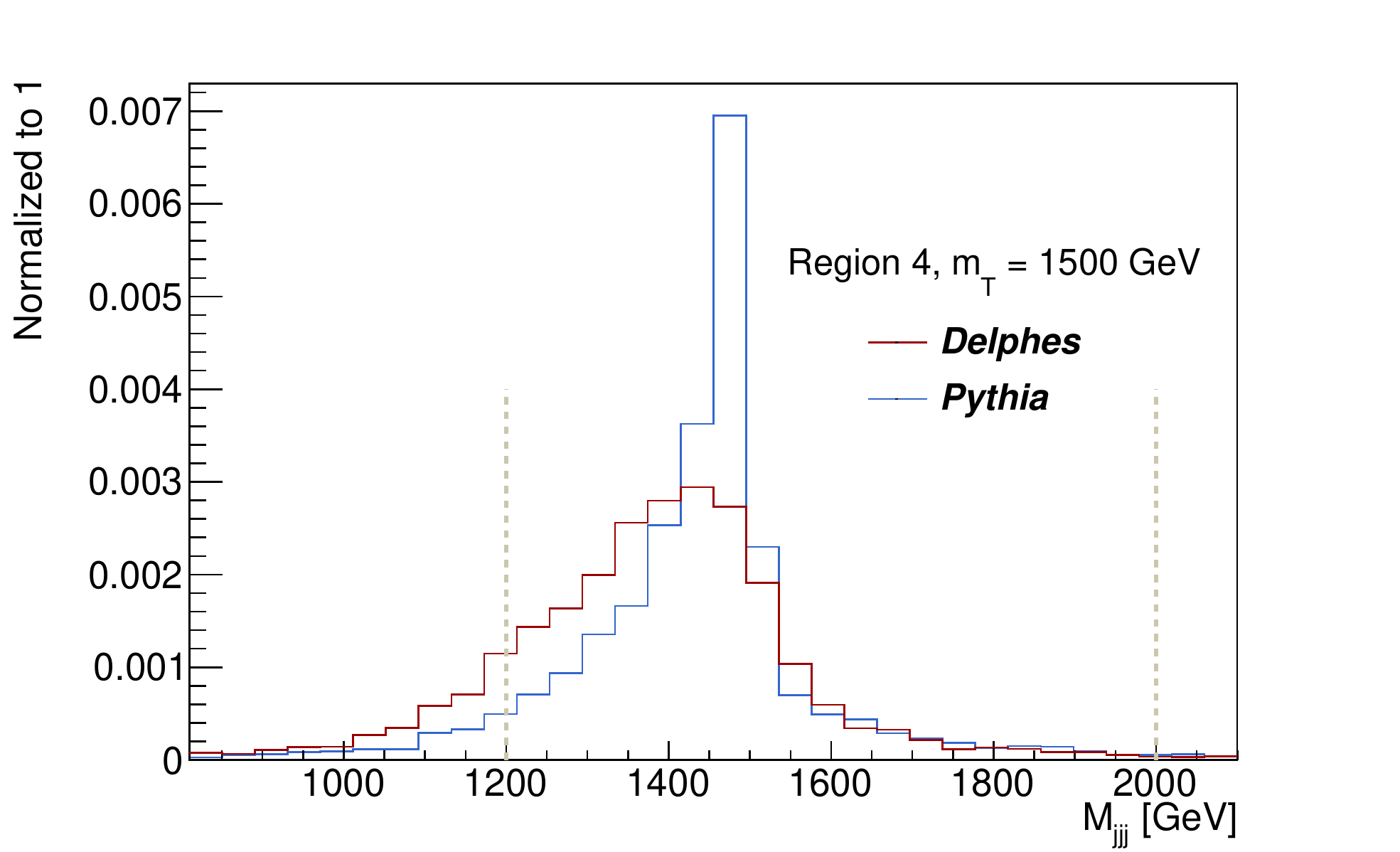}
\caption{  }
\end{subfigure}

\caption{\label{fig:CMS_mjjjDistribution} Mass distributions for two of the mass regions. The distributions in figure (a) and (b) are produced by CMS \cite{Sirunyan:2018duw} while the bottom figures are the corresponding ones for $m_{\eta} = 5 $ TeV. }
\end{figure}

Furthermore one of the most efficient cut for the three-jet resonance is the "Delta cut" defined as:
\begin{equation}
M_{jjj} < \sum_{i=1}^{3} p_{T}^{i} - \Delta  \; \;,
\end{equation}
where $M_{jjj}$ is the invariant mass of the triplet and $\Delta$ is an adjustable parameter. The parameter $\Delta$ is determined in each signal region by optimizing the signal significance $\alpha = s / \sqrt{s+b}$. This particular selection criteria can be understood due to the observation of the linear correlation of the triplet invariant mass with scalar sum of the transverse momentum for the QCD background. While the triplet invariant mass of the correctly combined signal triplets is unchanged by varying $p_{T}$ since $M_{jjj}$ is fixed. Consequently, this not only reduces the QCD multijet background but the combinatoric background raising from the incorrectly combined signal triplets as well. Finally, the mass asymmetry variable is defined as:
\begin{equation}
A_{m} = \frac{ \abs{ m_{ijk} - m_{lmn} } }{ m_{ijk} + m_{lmn} } \; \;,
\end{equation}
where $m_{ijk}$ is invariant mass of the triplet. This variable has discriminating power between signal and background since the signal triplets are expected to be close each other in mass but not the background.

When trying to reproduce the 13 TeV CMS search~\cite{Sirunyan:2018duw} we encountered some difficulites. The CMS collaboration paper contains $M_{jjj}$ distributions for their signal topologies reproduced in the top panels of figure~\ref{fig:CMS_mjjjDistribution}. The shapes of the signals appear as perfect Gaussians centered around the gluino mass. However, our simulations of the RPV model result in an invariant mass peak that is slightly shifted below the true mass points and is asymmetric about the peak with a longer tail at lower invariant mass. We show the signal distributions for two different gluino masses in the bottom panel of figure~\ref{fig:CMS_mjjjDistribution} both with and without detector simulation.
 Furthermore, it is stated in~\cite{Sirunyan:2018duw} that the invariant mass distribution of the incorrectly combined signal triplets (the combinatoric background) is similar to the multijet background; however, we find them to be different. There is also an ambiguity in the definition of the acceptance in the case where more than two triplets in an event satisfy all the selection criteria. Finally, \cite{Sirunyan:2018duw} refers to Monte Carlo simulations of the QCD background, but the work does not specify how the QCD samples are generated. More details about the procedures for computing the signal efficiency and simulating the background would be helpful for future studies and recasts.

\section{ATLAS Di-Jet Resonance Search}
\label{app:ATLAS_diJet_search}
Similar to the three-jet CMS search, this di-jet search~\cite{Aaboud:2017nmi} also reconstructs the jet candidates using an anti-$k_{t}$ algorithm with a radius parameter of 0.4. The complete list of cuts is displayed in table~\ref{table:ATLAS_selectionTable}. The average mass of the two reconstructed resonances is expected to peak around the mass of the resonance being searched for. The average mass,
\begin{equation}
m_{\text{avg}} = \frac{1}{2} \left( m_{1} + m_{2} \right)  \; \;,
\end{equation}
is thus required to be inside of a window around the searched for mass, with the width of the window varying from 10 to 100 GeV and is given in tables 3 and 5 of~\cite{Aaboud:2017nmi}. 
 In order to recast this particular search, the RPV top squarks were pair produced with radiation of up to two additional partons. The merging with parton shower was done using the MLM \cite{Alwall:2007fs} prescription with a merging scale set to $1/4$ of the top squark mass. In addition, all the SUSY particles except top squark were decoupled by setting their masses to 5 TeV.

\begin{table}[h]
\caption{The list of selection criteria with the direction of the cut for the ATLAS di-jet search~\cite{Aaboud:2017nmi}.}
\centering
\begin{tabular}{c c c c c }
\hline\hline
Jet $p_{T}$ [GeV] &   $\mathcal{A}_{m} $ & $\abs{ \cos (\theta^{\ast})}$ & $ \Delta R_{min} $ \\ [0.5ex] 
\hline
 \multirow{2}{*}{ $> 120$ } &  \multirow{2}{*}{ $< 0.05$ }  & \multirow{2}{*}{ $<0.3$ }  & $< -0.002 \cdot ( \frac{m_{avg} }{GeV} - 225 ) + 0.72  \ , \quad \text{if } m_{avg} \leq 225 \text{ GeV}$ \\
 	& & & $< +0.0013 \cdot ( \frac{m_{avg} }{GeV} - 225 ) + 0.72  \ , \quad \text{if } m_{avg} > 225 \text{ GeV}$ \\
\hline
\end{tabular}
\label{table:ATLAS_selectionTable}
\end{table}

\section{CMS Three Jet Resonance Search at $\sqrt{s} = 8 $ TeV}
\label{app:CMS_threeJet_8TeV_search}
The work of~\cite{Chatrchyan:2013gia} is an earlier version of the three-jet CMS search \cite{Sirunyan:2018duw} performed at the center-of-mass energy of 8 TeV. The jet candidates are constructed using an anti-$k_{t}$ algorithm with a radius parameter of 0.5. The search considers two scenarios, first when the gluino decays into light flavour jets and secondly when it decays to a $b$-jet and two light flavour jets. The latter case requires the existence of at least one bottom quark jet in the resonance decay products. Besides the usual $p_{T}$ and $\Delta$ variables requirements described in appendix~\ref{app:cms_threeJet_search}, event shape information is exploited. Typically in the high mass region, the signal events have a more spherical shape than the background (which generally contain back to back jets thus more linear shape)~\cite{Chatrchyan:2013gia}. Consequently, the sphericity variable is defined as,
\begin{equation}
S = \frac{3}{2} \left ( \lambda_{2} + \lambda_{3} \right )  \ , \quad \lambda_{1} \geq \lambda_{2} \geq \lambda_{3}  \; \;,
\end{equation} 
where $\lambda$'s are the eigenvalues of the sphericity tensor,
\begin{equation}
S^{\alpha \beta} = \frac{\sum \limits_{i} p_{i}^{\alpha} p_{i}^{\beta}  }{\sum \limits_{i} \abs{p_{i}}^{2}} \ , \quad \alpha \ ,\  \beta = \text{ x , y , z} \; \;,
\end{equation}
where $\alpha$ and $\beta$ label separate jets, and the sphericity $S$ is calculated using all jets in each event. The complete list of selection criteria are shown in table~\ref{table:CMS_8TeV_selectionTable}.

\begin{table}[h]
\caption{The selection criteria with the direction of the cut for the CMS heavy flavour search performed at $\sqrt{s} = 8$ TeV~\cite{Chatrchyan:2013gia}.}
\centering
\begin{tabular}{c c c c c c}
\hline\hline
Mass Range [GeV] & $\Delta$ [GeV]  & $p_{T,j}^{4th} $ [GeV] & $p_{T,j}^{6th} $ [GeV] & Sphericity \\ [0.5ex] 
\hline 
200-600 	& $> 110$   & $> 80$ & $> 60$ & --- \\
600-1500 	& $> 110$   &  $>110$ & $> 110$ & $> 0.4$  \\
\hline
\end{tabular}
\label{table:CMS_8TeV_selectionTable}
\end{table}

\section{Multijet Background}
\label{app:QCD_section}
The principal background for our signal arises from the QCD multijet events. Other SM processes have negligible contributions, and we have performed simulations of $\bar{t} t$ events to confirm that their rates are indeed very small. The QCD multijet background is very large as one can observe from the crude cross section estimates shown in table~\ref{table:QCD_Xsection} (similar results were obtained using Sherpa~\cite{Gleisberg:2008ta}). The QCD multijet events were obtained by simulating $pp\rightarrow j j j j$ using MadGraph5 interfaced with Pythia8 and Delphes. The cross sections are orders of magnitude larger than the pair production cross section for top partner as displayed in figure~\ref{fig:myCrossection}. In our simulations, we require each of the four partons to have $p_T > p_{T,min} (j) = 100 $ GeV in order to make sure enough events satisfy all the selection requirements for our recasted searches. The minimum parton level $p_{T}$ is well below the detector level jet $p_{T} $ requirement of 125 GeV, given in table~\ref{table:CMS_selectionTable}, at the cut off region ($m_{T} = 900$ GeV), so it does not affect our analysis.  

\begin{table}[h]
\caption{ Four partonic hard jets production cross section using MadGraph5 at $\sqrt{s} = 13 $ TeV with various minimum parton level cut. }
\centering
\begin{tabular}{ c c } 
\hline
$p_{T,min} (j)$ Generator Level [GeV]& $\sigma_{4j}$ [pb]  \\
\hline
20 & $1.79 \times 10^{7} $  \\
60 & $7.64 \times 10^{4} $    \\
100 & $4.68 \times 10^{3} $  \\
200 & $7.216 \times 10^{1} $  \\
\hline
\end{tabular}
\label{table:QCD_Xsection}
\end{table}

\newpage  
\bibliographystyle{JHEP}
\bibliography{references}

\providecommand{\href}[2]{#2}\begingroup\raggedright\begin{thebibliography}{10}

\bibitem{Aad:2012tfa}
{\scshape ATLAS} collaboration, \emph{{Observation of a new particle in the
  search for the Standard Model Higgs boson with the ATLAS detector at the
  LHC}}, \href{https://doi.org/10.1016/j.physletb.2012.08.020}{\emph{Phys.
  Lett.} {\bfseries B716} (2012) 1}
  [\href{https://arxiv.org/abs/1207.7214}{{\ttfamily 1207.7214}}].

\bibitem{Chatrchyan:2012xdj}
{\scshape CMS} collaboration, \emph{{Observation of a New Boson at a Mass of
  125 GeV with the CMS Experiment at the LHC}},
  \href{https://doi.org/10.1016/j.physletb.2012.08.021}{\emph{Phys. Lett.}
  {\bfseries B716} (2012) 30}
  [\href{https://arxiv.org/abs/1207.7235}{{\ttfamily 1207.7235}}].

\bibitem{Contino:2006qr}
R.~Contino, L.~Da~Rold and A.~Pomarol, \emph{{Light custodians in natural
  composite Higgs models}},
  \href{https://doi.org/10.1103/PhysRevD.75.055014}{\emph{Phys. Rev.}
  {\bfseries D75} (2007) 055014}
  [\href{https://arxiv.org/abs/hep-ph/0612048}{{\ttfamily hep-ph/0612048}}].

\bibitem{Marzocca:2012zn}
D.~Marzocca, M.~Serone and J.~Shu, \emph{{General Composite Higgs Models}},
  \href{https://doi.org/10.1007/JHEP08(2012)013}{\emph{JHEP} {\bfseries 08}
  (2012) 013} [\href{https://arxiv.org/abs/1205.0770}{{\ttfamily 1205.0770}}].

\bibitem{Matsedonskyi:2012ym}
O.~Matsedonskyi, G.~Panico and A.~Wulzer, \emph{{Light Top Partners for a Light
  Composite Higgs}}, \href{https://doi.org/10.1007/JHEP01(2013)164}{\emph{JHEP}
  {\bfseries 01} (2013) 164} [\href{https://arxiv.org/abs/1204.6333}{{\ttfamily
  1204.6333}}].

\bibitem{DeSimone:2012fs}
A.~De~Simone, O.~Matsedonskyi, R.~Rattazzi and A.~Wulzer, \emph{{A First Top
  Partner Hunter's Guide}},
  \href{https://doi.org/10.1007/JHEP04(2013)004}{\emph{JHEP} {\bfseries 04}
  (2013) 004} [\href{https://arxiv.org/abs/1211.5663}{{\ttfamily 1211.5663}}].

\bibitem{Bellazzini:2014yua}
B.~Bellazzini, C.~Csáki and J.~Serra, \emph{{Composite Higgses}},
  \href{https://doi.org/10.1140/epjc/s10052-014-2766-x}{\emph{Eur. Phys. J.}
  {\bfseries C74} (2014) 2766}
  [\href{https://arxiv.org/abs/1401.2457}{{\ttfamily 1401.2457}}].

\bibitem{ArkaniHamed:2002qy}
N.~Arkani-Hamed, A.~G. Cohen, E.~Katz and A.~E. Nelson, \emph{{The Littlest
  Higgs}}, \href{https://doi.org/10.1088/1126-6708/2002/07/034}{\emph{JHEP}
  {\bfseries 07} (2002) 034}
  [\href{https://arxiv.org/abs/hep-ph/0206021}{{\ttfamily hep-ph/0206021}}].

\bibitem{Schmaltz:2005ky}
M.~Schmaltz and D.~Tucker-Smith, \emph{{Little Higgs review}},
  \href{https://doi.org/10.1146/annurev.nucl.55.090704.151502}{\emph{Ann. Rev.
  Nucl. Part. Sci.} {\bfseries 55} (2005) 229}
  [\href{https://arxiv.org/abs/hep-ph/0502182}{{\ttfamily hep-ph/0502182}}].

\bibitem{Aaboud:2017qpr}
{\scshape ATLAS} collaboration, \emph{{Search for pair production of
  vector-like top quarks in events with one lepton, jets, and missing
  transverse momentum in $ \sqrt{s}=13 $ TeV $pp$ collisions with the ATLAS
  detector}}, \href{https://doi.org/10.1007/JHEP08(2017)052}{\emph{JHEP}
  {\bfseries 08} (2017) 052}
  [\href{https://arxiv.org/abs/1705.10751}{{\ttfamily 1705.10751}}].

\bibitem{Sirunyan:2017usq}
{\scshape CMS} collaboration, \emph{{Search for pair production of vector-like
  T and B quarks in single-lepton final states using boosted jet substructure
  in proton-proton collisions at $\sqrt{s}=13$ TeV}},
  \href{https://doi.org/10.1007/JHEP11(2017)085}{\emph{JHEP} {\bfseries 11}
  (2017) 085} [\href{https://arxiv.org/abs/1706.03408}{{\ttfamily
  1706.03408}}].

\bibitem{Sirunyan:2017pks}
{\scshape CMS} collaboration, \emph{{Search for pair production of vector-like
  quarks in the bW$\overline{\mathrm{b}}$W channel from proton-proton
  collisions at $\sqrt{s} =$ 13 TeV}},
  \href{https://doi.org/10.1016/j.physletb.2018.01.077}{\emph{Phys. Lett.}
  {\bfseries B779} (2018) 82}
  [\href{https://arxiv.org/abs/1710.01539}{{\ttfamily 1710.01539}}].

\bibitem{Aaboud:2018xuw}
{\scshape ATLAS} collaboration, \emph{{Search for pair production of up-type
  vector-like quarks and for four-top-quark events in final states with
  multiple $b$-jets with the ATLAS detector}},
  \href{https://doi.org/10.1007/JHEP07(2018)089}{\emph{JHEP} {\bfseries 07}
  (2018) 089} [\href{https://arxiv.org/abs/1803.09678}{{\ttfamily
  1803.09678}}].

\bibitem{Sirunyan:2018omb}
{\scshape CMS} collaboration, \emph{{Search for vector-like T and B quark pairs
  in final states with leptons at $\sqrt{s} =$ 13 TeV}},
  \href{https://doi.org/10.1007/JHEP08(2018)177}{\emph{JHEP} {\bfseries 08}
  (2018) 177} [\href{https://arxiv.org/abs/1805.04758}{{\ttfamily
  1805.04758}}].

\bibitem{Aaboud:2018uek}
{\scshape ATLAS} collaboration, \emph{{Search for pair production of heavy
  vector-like quarks decaying into high-$p_T$ $W$ bosons and top quarks in the
  lepton-plus-jets final state in $pp$ collisions at $\sqrt{s}=13$ TeV with the
  ATLAS detector}}, \href{https://doi.org/10.1007/JHEP08(2018)048}{\emph{JHEP}
  {\bfseries 08} (2018) 048}
  [\href{https://arxiv.org/abs/1806.01762}{{\ttfamily 1806.01762}}].

\bibitem{Aaboud:2018wxv}
{\scshape ATLAS} collaboration, \emph{{Search for pair production of heavy
  vector-like quarks decaying into hadronic final states in $pp$ collisions at
  $\sqrt{s} = 13$ TeV with the ATLAS detector}},
  \href{https://doi.org/10.1103/PhysRevD.98.092005}{\emph{Phys. Rev.}
  {\bfseries D98} (2018) 092005}
  [\href{https://arxiv.org/abs/1808.01771}{{\ttfamily 1808.01771}}].

\bibitem{Aaboud:2018pii}
{\scshape ATLAS} collaboration, \emph{{Combination of the searches for
  pair-produced vector-like partners of the third-generation quarks at
  $\sqrt{s} =$ 13 TeV with the ATLAS detector}},
  \href{https://doi.org/10.1103/PhysRevLett.121.211801}{\emph{Phys. Rev. Lett.}
  {\bfseries 121} (2018) 211801}
  [\href{https://arxiv.org/abs/1808.02343}{{\ttfamily 1808.02343}}].

\bibitem{Sirunyan:2018qau}
{\scshape CMS} collaboration, \emph{{Search for vector-like quarks in events
  with two oppositely charged leptons and jets in proton-proton collisions at
  $\sqrt{s} =$ 13 TeV}},
  \href{https://doi.org/10.1140/epjc/s10052-019-6855-8}{\emph{Eur. Phys. J.}
  {\bfseries C79} (2019) 364}
  [\href{https://arxiv.org/abs/1812.09768}{{\ttfamily 1812.09768}}].

\bibitem{Sirunyan:2019sza}
{\scshape CMS} collaboration, \emph{{Search for pair production of vector-like
  quarks in the fully hadronic final state}},
  \href{https://arxiv.org/abs/1906.11903}{{\ttfamily 1906.11903}}.

\bibitem{Aad:2014efa}
{\scshape ATLAS} collaboration, \emph{{Search for pair and single production of
  new heavy quarks that decay to a $Z$ boson and a third-generation quark in
  $pp$ collisions at $\sqrt{s}=8$ TeV with the ATLAS detector}},
  \href{https://doi.org/10.1007/JHEP11(2014)104}{\emph{JHEP} {\bfseries 11}
  (2014) 104} [\href{https://arxiv.org/abs/1409.5500}{{\ttfamily 1409.5500}}].

\bibitem{Aad:2015voa}
{\scshape ATLAS} collaboration, \emph{{Search for the production of single
  vector-like and excited quarks in the $Wt$ final state in $pp$ collisions at
  $\sqrt{s}$ = 8 TeV with the ATLAS detector}},
  \href{https://doi.org/10.1007/JHEP02(2016)110}{\emph{JHEP} {\bfseries 02}
  (2016) 110} [\href{https://arxiv.org/abs/1510.02664}{{\ttfamily
  1510.02664}}].

\bibitem{Aad:2016qpo}
{\scshape ATLAS} collaboration, \emph{{Search for single production of
  vector-like quarks decaying into Wb in pp collisions at $\sqrt{s} = 8$ TeV
  with the ATLAS detector}},
  \href{https://doi.org/10.1140/epjc/s10052-016-4281-8}{\emph{Eur. Phys. J.}
  {\bfseries C76} (2016) 442}
  [\href{https://arxiv.org/abs/1602.05606}{{\ttfamily 1602.05606}}].

\bibitem{Khachatryan:2016vph}
{\scshape CMS} collaboration, \emph{{Search for single production of a heavy
  vector-like T quark decaying to a Higgs boson and a top quark with a lepton
  and jets in the final state}},
  \href{https://doi.org/10.1016/j.physletb.2017.05.019}{\emph{Phys. Lett.}
  {\bfseries B771} (2017) 80}
  [\href{https://arxiv.org/abs/1612.00999}{{\ttfamily 1612.00999}}].

\bibitem{Sirunyan:2016ipo}
{\scshape CMS} collaboration, \emph{{Search for electroweak production of a
  vector-like quark decaying to a top quark and a Higgs boson using boosted
  topologies in fully hadronic final states}},
  \href{https://doi.org/10.1007/JHEP04(2017)136}{\emph{JHEP} {\bfseries 04}
  (2017) 136} [\href{https://arxiv.org/abs/1612.05336}{{\ttfamily
  1612.05336}}].

\bibitem{Sirunyan:2017ezy}
{\scshape CMS} collaboration, \emph{{Search for single production of
  vector-like quarks decaying to a Z boson and a top or a bottom quark in
  proton-proton collisions at $ \sqrt{s}=13 $ TeV}},
  \href{https://doi.org/10.1007/JHEP05(2017)029}{\emph{JHEP} {\bfseries 05}
  (2017) 029} [\href{https://arxiv.org/abs/1701.07409}{{\ttfamily
  1701.07409}}].

\bibitem{Sirunyan:2017tfc}
{\scshape CMS} collaboration, \emph{{Search for single production of
  vector-like quarks decaying into a b quark and a W boson in proton-proton
  collisions at $\sqrt s =$ 13 TeV}},
  \href{https://doi.org/10.1016/j.physletb.2017.07.022}{\emph{Phys. Lett.}
  {\bfseries B772} (2017) 634}
  [\href{https://arxiv.org/abs/1701.08328}{{\ttfamily 1701.08328}}].

\bibitem{Sirunyan:2017ynj}
{\scshape CMS} collaboration, \emph{{Search for single production of a
  vector-like T quark decaying to a Z boson and a top quark in proton-proton
  collisions at $\sqrt s$ = 13 TeV}},
  \href{https://doi.org/10.1016/j.physletb.2018.04.036}{\emph{Phys. Lett.}
  {\bfseries B781} (2018) 574}
  [\href{https://arxiv.org/abs/1708.01062}{{\ttfamily 1708.01062}}].

\bibitem{Sirunyan:2018fjh}
{\scshape CMS} collaboration, \emph{{Search for single production of
  vector-like quarks decaying to a b quark and a Higgs boson}},
  \href{https://doi.org/10.1007/JHEP06(2018)031}{\emph{JHEP} {\bfseries 06}
  (2018) 031} [\href{https://arxiv.org/abs/1802.01486}{{\ttfamily
  1802.01486}}].

\bibitem{Aaboud:2018saj}
{\scshape ATLAS} collaboration, \emph{{Search for pair- and single-production
  of vector-like quarks in final states with at least one $Z$ boson decaying
  into a pair of electrons or muons in $pp$ collision data collected with the
  ATLAS detector at $\sqrt{s} = 13$ TeV}},
  \href{https://doi.org/10.1103/PhysRevD.98.112010}{\emph{Phys. Rev.}
  {\bfseries D98} (2018) 112010}
  [\href{https://arxiv.org/abs/1806.10555}{{\ttfamily 1806.10555}}].

\bibitem{Sirunyan:2018ncp}
{\scshape CMS} collaboration, \emph{{Search for single production of
  vector-like quarks decaying to a top quark and a W boson in proton-proton
  collisions at $\sqrt{s} =$ 13 TeV}},
  \href{https://doi.org/10.1140/epjc/s10052-019-6556-3}{\emph{Eur. Phys. J. C}
  {\bfseries 79} (2019) 90} [\href{https://arxiv.org/abs/1809.08597}{{\ttfamily
  1809.08597}}].

\bibitem{Anandakrishnan:2015yfa}
A.~Anandakrishnan, J.~H. Collins, M.~Farina, E.~Kuflik and M.~Perelstein,
  \emph{{Odd Top Partners at the LHC}},
  \href{https://doi.org/10.1103/PhysRevD.93.075009}{\emph{Phys. Rev.}
  {\bfseries D93} (2016) 075009}
  [\href{https://arxiv.org/abs/1506.05130}{{\ttfamily 1506.05130}}].

\bibitem{Aguilar-Saavedra:2017giu}
J.~A. Aguilar-Saavedra, D.~E. López-Fogliani and C.~Muñoz, \emph{{Novel
  signatures for vector-like quarks}},
  \href{https://doi.org/10.1007/JHEP06(2017)095}{\emph{JHEP} {\bfseries 06}
  (2017) 095} [\href{https://arxiv.org/abs/1705.02526}{{\ttfamily
  1705.02526}}].

\bibitem{Aguilar-Saavedra:2019ghg}
J.~A. Aguilar-Saavedra, J.~Alonso-González, L.~Merlo and J.~M. No,
  \emph{{Exotic vectorlike quark phenomenology in the minimal linear $\sigma$
  model}}, \href{https://doi.org/10.1103/PhysRevD.101.035015}{\emph{Phys. Rev.}
  {\bfseries D101} (2020) 035015}
  [\href{https://arxiv.org/abs/1911.10202}{{\ttfamily 1911.10202}}].

\bibitem{Kim:2018mks}
J.~H. Kim and I.~M. Lewis, \emph{{Loop Induced Single Top Partner Production
  and Decay at the LHC}},
  \href{https://doi.org/10.1007/JHEP05(2018)095}{\emph{JHEP} {\bfseries 05}
  (2018) 095} [\href{https://arxiv.org/abs/1803.06351}{{\ttfamily
  1803.06351}}].

\bibitem{Alhazmi:2018whk}
H.~Alhazmi, J.~H. Kim, K.~Kong and I.~M. Lewis, \emph{{Shedding Light on Top
  Partner at the LHC}},
  \href{https://doi.org/10.1007/JHEP01(2019)139}{\emph{JHEP} {\bfseries 01}
  (2019) 139} [\href{https://arxiv.org/abs/1808.03649}{{\ttfamily
  1808.03649}}].

\bibitem{Benbrik:2019zdp}
R.~Benbrik et~al., \emph{{Signatures of vector-like top partners decaying into
  new neutral scalar or pseudoscalar bosons}},
  \href{https://arxiv.org/abs/1907.05929}{{\ttfamily 1907.05929}}.

\bibitem{Kim:2019oyh}
J.~H. Kim, S.~D. Lane, H.-S. Lee, I.~M. Lewis and M.~Sullivan, \emph{{Searching
  for Dark Photons with Maverick Top Partners}},
  \href{https://arxiv.org/abs/1904.05893}{{\ttfamily 1904.05893}}.

\bibitem{Chala:2017xgc}
M.~Chala, \emph{{Direct bounds on heavy toplike quarks with standard and exotic
  decays}}, \href{https://doi.org/10.1103/PhysRevD.96.015028}{\emph{Phys. Rev.}
  {\bfseries D96} (2017) 015028}
  [\href{https://arxiv.org/abs/1705.03013}{{\ttfamily 1705.03013}}].

\bibitem{Bizot:2018tds}
N.~Bizot, G.~Cacciapaglia and T.~Flacke, \emph{{Common exotic decays of top
  partners}}, \href{https://doi.org/10.1007/JHEP06(2018)065}{\emph{JHEP}
  {\bfseries 06} (2018) 065}
  [\href{https://arxiv.org/abs/1803.00021}{{\ttfamily 1803.00021}}].

\bibitem{Cacciapaglia:2019zmj}
G.~Cacciapaglia, T.~Flacke, M.~Park and M.~Zhang, \emph{{Exotic decays of top
  partners: mind the search gap}},
  \href{https://arxiv.org/abs/1908.07524}{{\ttfamily 1908.07524}}.

\bibitem{TheATLAScollaboration:2015atd}
\emph{{Performance and Calibration of the JetFitterCharm Algorithm for c-Jet
  Identification ATL-PHYS-PUB-2015-001, 2015}}, .

\bibitem{CMS:2016knj}
{\scshape CMS} collaboration, \emph{{Identification of c-quark jets at the CMS
  experiment CMS-PAS-BTV-16-001, 2016}}, .

\bibitem{Sirunyan:2017ezt}
{\scshape CMS} collaboration, \emph{{Identification of heavy-flavour jets with
  the CMS detector in pp collisions at 13 TeV}},
  \href{https://doi.org/10.1088/1748-0221/13/05/P05011}{\emph{JINST} {\bfseries
  13} (2018) P05011} [\href{https://arxiv.org/abs/1712.07158}{{\ttfamily
  1712.07158}}].

\bibitem{Aaboud:2018xwy}
{\scshape ATLAS} collaboration, \emph{{Measurements of b-jet tagging efficiency
  with the ATLAS detector using $ t\overline{t} $ events at $ \sqrt{s}=13 $
  TeV}}, \href{https://doi.org/10.1007/JHEP08(2018)089}{\emph{JHEP} {\bfseries
  08} (2018) 089} [\href{https://arxiv.org/abs/1805.01845}{{\ttfamily
  1805.01845}}].

\bibitem{Kats:2017ojr}
Y.~Kats, M.~McCullough, G.~Perez, Y.~Soreq and J.~Thaler, \emph{{Colorful
  Twisted Top Partners and Partnerium at the LHC}},
  \href{https://doi.org/10.1007/JHEP06(2017)126}{\emph{JHEP} {\bfseries 06}
  (2017) 126} [\href{https://arxiv.org/abs/1704.03393}{{\ttfamily
  1704.03393}}].

\bibitem{Dreiner:1997uz}
H.~K. Dreiner, \emph{{An Introduction to explicit R-parity violation}},
  \href{https://arxiv.org/abs/hep-ph/9707435}{{\ttfamily hep-ph/9707435}}.

\bibitem{Allanach:2003eb}
B.~C. Allanach, A.~Dedes and H.~K. Dreiner, \emph{{R parity violating minimal
  supergravity model}}, \href{https://doi.org/10.1103/PhysRevD.69.115002,
  10.1103/PhysRevD.72.079902}{\emph{Phys. Rev.} {\bfseries D69} (2004) 115002}
  [\href{https://arxiv.org/abs/hep-ph/0309196}{{\ttfamily hep-ph/0309196}}].

\bibitem{Barbier:2004ez}
R.~Barbier et~al., \emph{{R-parity violating supersymmetry}},
  \href{https://doi.org/10.1016/j.physrep.2005.08.006}{\emph{Phys. Rept.}
  {\bfseries 420} (2005) 1}
  [\href{https://arxiv.org/abs/hep-ph/0406039}{{\ttfamily hep-ph/0406039}}].

\bibitem{Sirunyan:2018duw}
{\scshape CMS} collaboration, \emph{{Search for pair-produced three-jet
  resonances in proton-proton collisions at $\sqrt s$ =13 TeV}},
  \href{https://doi.org/10.1103/PhysRevD.99.012010}{\emph{Phys. Rev.}
  {\bfseries D99} (2019) 012010}
  [\href{https://arxiv.org/abs/1810.10092}{{\ttfamily 1810.10092}}].

\bibitem{Aaboud:2017nmi}
{\scshape ATLAS} collaboration, \emph{{A search for pair-produced resonances in
  four-jet final states at $\sqrt{s} =$ 13 TeV with the ATLAS detector}},
  \href{https://doi.org/10.1140/epjc/s10052-018-5693-4}{\emph{Eur. Phys. J.}
  {\bfseries C78} (2018) 250}
  [\href{https://arxiv.org/abs/1710.07171}{{\ttfamily 1710.07171}}].

\bibitem{Alloul:2013bka}
A.~Alloul, N.~D. Christensen, C.~Degrande, C.~Duhr and B.~Fuks,
  \emph{{FeynRules 2.0 - A complete toolbox for tree-level phenomenology}},
  \href{https://doi.org/10.1016/j.cpc.2014.04.012}{\emph{Comput. Phys. Commun.}
  {\bfseries 185} (2014) 2250}
  [\href{https://arxiv.org/abs/1310.1921}{{\ttfamily 1310.1921}}].

\bibitem{Alwall:2014hca}
J.~Alwall, R.~Frederix, S.~Frixione, V.~Hirschi, F.~Maltoni, O.~Mattelaer
  et~al., \emph{{The automated computation of tree-level and next-to-leading
  order differential cross sections, and their matching to parton shower
  simulations}}, \href{https://doi.org/10.1007/JHEP07(2014)079}{\emph{JHEP}
  {\bfseries 07} (2014) 079} [\href{https://arxiv.org/abs/1405.0301}{{\ttfamily
  1405.0301}}].

\bibitem{Sjostrand:2014zea}
T.~Sjöstrand, S.~Ask, J.~R. Christiansen, R.~Corke, N.~Desai, P.~Ilten et~al.,
  \emph{{An Introduction to PYTHIA 8.2}},
  \href{https://doi.org/10.1016/j.cpc.2015.01.024}{\emph{Comput. Phys. Commun.}
  {\bfseries 191} (2015) 159}
  [\href{https://arxiv.org/abs/1410.3012}{{\ttfamily 1410.3012}}].

\bibitem{deFavereau:2013fsa}
{\scshape DELPHES 3} collaboration, \emph{{DELPHES 3, A modular framework for
  fast simulation of a generic collider experiment}},
  \href{https://doi.org/10.1007/JHEP02(2014)057}{\emph{JHEP} {\bfseries 02}
  (2014) 057} [\href{https://arxiv.org/abs/1307.6346}{{\ttfamily 1307.6346}}].

\bibitem{Cacciari:2011ma}
M.~Cacciari, G.~P. Salam and G.~Soyez, \emph{{FastJet User Manual}},
  \href{https://doi.org/10.1140/epjc/s10052-012-1896-2}{\emph{Eur. Phys. J.}
  {\bfseries C72} (2012) 1896}
  [\href{https://arxiv.org/abs/1111.6097}{{\ttfamily 1111.6097}}].

\bibitem{Czakon:2011xx}
M.~Czakon and A.~Mitov, \emph{{Top++: A Program for the Calculation of the
  Top-Pair Cross-Section at Hadron Colliders}},
  \href{https://doi.org/10.1016/j.cpc.2014.06.021}{\emph{Comput. Phys. Commun.}
  {\bfseries 185} (2014) 2930}
  [\href{https://arxiv.org/abs/1112.5675}{{\ttfamily 1112.5675}}].

\bibitem{Chatrchyan:2012uxa}
{\scshape CMS} collaboration, \emph{{Search for Three-Jet Resonances in $pp$
  Collisions at $\sqrt{s}=7$ TeV}},
  \href{https://doi.org/10.1016/j.physletb.2012.10.048}{\emph{Phys. Lett.}
  {\bfseries B718} (2012) 329}
  [\href{https://arxiv.org/abs/1208.2931}{{\ttfamily 1208.2931}}].

\bibitem{ATLAS:2012dp}
{\scshape ATLAS} collaboration, \emph{{Search for pair production of massive
  particles decaying into three quarks with the ATLAS detector in $\sqrt{s}=7$
  TeV $pp$ collisions at the LHC}},
  \href{https://doi.org/10.1007/JHEP12(2012)086}{\emph{JHEP} {\bfseries 12}
  (2012) 086} [\href{https://arxiv.org/abs/1210.4813}{{\ttfamily 1210.4813}}].

\bibitem{Aad:2015lea}
{\scshape ATLAS} collaboration, \emph{{Search for massive supersymmetric
  particles decaying to many jets using the ATLAS detector in $pp$ collisions
  at $\sqrt{s} = 8$ TeV}}, \href{https://doi.org/10.1103/PhysRevD.93.039901,
  10.1103/PhysRevD.91.112016}{\emph{Phys. Rev.} {\bfseries D91} (2015) 112016}
  [\href{https://arxiv.org/abs/1502.05686}{{\ttfamily 1502.05686}}].

\bibitem{Redi:2013eaa}
M.~Redi, V.~Sanz, M.~de~Vries and A.~Weiler, \emph{{Strong Signatures of
  Right-Handed Compositeness}},
  \href{https://doi.org/10.1007/JHEP08(2013)008}{\emph{JHEP} {\bfseries 08}
  (2013) 008} [\href{https://arxiv.org/abs/1305.3818}{{\ttfamily 1305.3818}}].

\bibitem{Dalitz:1954cq}
R.~H. Dalitz, \emph{{Decay of tau mesons of known charge}},
  \href{https://doi.org/10.1103/PhysRev.94.1046}{\emph{Phys. Rev.} {\bfseries
  94} (1954) 1046}.

\bibitem{Junk:1999kv}
T.~Junk, \emph{{Confidence level computation for combining searches with small
  statistics}},
  \href{https://doi.org/10.1016/S0168-9002(99)00498-2}{\emph{Nucl. Instrum.
  Meth.} {\bfseries A434} (1999) 435}
  [\href{https://arxiv.org/abs/hep-ex/9902006}{{\ttfamily hep-ex/9902006}}].

\bibitem{Read_2002}
A.~L. Read, \emph{Presentation of search results: {theCLstechnique}},
  \href{https://doi.org/10.1088/0954-3899/28/10/313}{\emph{Journal of Physics
  G: Nuclear and Particle Physics} {\bfseries 28} (2002) 2693}.

\bibitem{Cowan:2010js}
G.~Cowan, K.~Cranmer, E.~Gross and O.~Vitells, \emph{{Asymptotic formulae for
  likelihood-based tests of new physics}},
  \href{https://doi.org/10.1140/epjc/s10052-011-1554-0,
  10.1140/epjc/s10052-013-2501-z}{\emph{Eur. Phys. J.} {\bfseries C71} (2011)
  1554} [\href{https://arxiv.org/abs/1007.1727}{{\ttfamily 1007.1727}}].

\bibitem{Aaltonen:2011sg}
{\scshape CDF} collaboration, \emph{{First Search for Multijet Resonances in
  $\sqrt{s} = 1.96$ TeV $ p\bar{p}$ Collisions}},
  \href{https://doi.org/10.1103/PhysRevLett.107.042001}{\emph{Phys. Rev. Lett.}
  {\bfseries 107} (2011) 042001}
  [\href{https://arxiv.org/abs/1105.2815}{{\ttfamily 1105.2815}}].

\bibitem{Barate:1999fc}
{\scshape ALEPH} collaboration, \emph{{Search for R-parity violating decays of
  supersymmetric particles in $e^{+} e^{-}$ collisions at center-of-mass
  energies near 183-GeV}},
  \href{https://doi.org/10.1007/s100520000284}{\emph{Eur. Phys. J. C}
  {\bfseries 13} (2000) 29}.

\bibitem{Heister:2002jc}
{\scshape ALEPH} collaboration, \emph{{Search for supersymmetric particles with
  R parity violating decays in $e^{+} e^{-}$ collisions at $\sqrt{s}$ up to
  209-GeV}}, \href{https://doi.org/10.1140/epjc/s2003-01311-5}{\emph{Eur. Phys.
  J. C} {\bfseries 31} (2003) 1}
  [\href{https://arxiv.org/abs/hep-ex/0210014}{{\ttfamily hep-ex/0210014}}].

\bibitem{Chatrchyan:2013ual}
{\scshape CMS} collaboration, \emph{{Measurement of the $t\bar{t}$ Production
  Cross Section in the All-Jet Final State in pp Collisions at $\sqrt{s}$ = 7
  TeV}}, \href{https://doi.org/10.1007/JHEP05(2013)065}{\emph{JHEP} {\bfseries
  05} (2013) 065} [\href{https://arxiv.org/abs/1302.0508}{{\ttfamily
  1302.0508}}].

\bibitem{Aad:2014zea}
{\scshape ATLAS} collaboration, \emph{{Measurement of the top-quark mass in the
  fully hadronic decay channel from ATLAS data at $\sqrt{s}=7\mathrm{\,TeV}$}},
  \href{https://doi.org/10.1140/epjc/s10052-015-3373-1}{\emph{Eur. Phys. J. C}
  {\bfseries 75} (2015) 158} [\href{https://arxiv.org/abs/1409.0832}{{\ttfamily
  1409.0832}}].

\bibitem{Khachatryan:2015fwh}
{\scshape CMS} collaboration, \emph{{Measurement of the
  $\mathrm{t}\overline{{\mathrm{t}}}$ production cross section in the all-jets
  final state in pp collisions at $\sqrt{s}=8$ $\,\text {TeV}$}},
  \href{https://doi.org/10.1140/epjc/s10052-016-3956-5}{\emph{Eur. Phys. J. C}
  {\bfseries 76} (2016) 128}
  [\href{https://arxiv.org/abs/1509.06076}{{\ttfamily 1509.06076}}].

\bibitem{Aaboud:2017mae}
{\scshape ATLAS} collaboration, \emph{{Top-quark mass measurement in the
  all-hadronic $ t\overline{t} $ decay channel at $ \sqrt{s}=8 $ TeV with the
  ATLAS detector}}, \href{https://doi.org/10.1007/JHEP09(2017)118}{\emph{JHEP}
  {\bfseries 09} (2017) 118}
  [\href{https://arxiv.org/abs/1702.07546}{{\ttfamily 1702.07546}}].

\bibitem{ATLAS:2020aub}
{\scshape ATLAS} collaboration, \emph{{Measurements of top-quark pair single-
  and double-differential cross-sections in the all-hadronic channel in $pp$
  collisions at $\sqrt{s}$ = 13 TeV using the ATLAS detector}}, .

\bibitem{Aaboud:2018lpl}
{\scshape ATLAS} collaboration, \emph{{Search for R-parity-violating
  supersymmetric particles in multi-jet final states produced in $p$-$p$
  collisions at $\sqrt{s} =13$ TeV using the ATLAS detector at the LHC}},
  \href{https://doi.org/10.1016/j.physletb.2018.08.021}{\emph{Phys. Lett.}
  {\bfseries B785} (2018) 136}
  [\href{https://arxiv.org/abs/1804.03568}{{\ttfamily 1804.03568}}].

\bibitem{Chatrchyan:2013gia}
{\scshape CMS} collaboration, \emph{{Searches for light- and heavy-flavour
  three-jet resonances in pp collisions at $\sqrt{s} = 8$ TeV}},
  \href{https://doi.org/10.1016/j.physletb.2014.01.049}{\emph{Phys. Lett.}
  {\bfseries B730} (2014) 193}
  [\href{https://arxiv.org/abs/1311.1799}{{\ttfamily 1311.1799}}].

\bibitem{Cacciari:2008gp}
M.~Cacciari, G.~P. Salam and G.~Soyez, \emph{{The anti-$k_t$ jet clustering
  algorithm}}, \href{https://doi.org/10.1088/1126-6708/2008/04/063}{\emph{JHEP}
  {\bfseries 04} (2008) 063} [\href{https://arxiv.org/abs/0802.1189}{{\ttfamily
  0802.1189}}].

\bibitem{Alwall:2007fs}
J.~Alwall et~al., \emph{{Comparative study of various algorithms for the
  merging of parton showers and matrix elements in hadronic collisions}},
  \href{https://doi.org/10.1140/epjc/s10052-007-0490-5}{\emph{Eur. Phys. J.}
  {\bfseries C53} (2008) 473}
  [\href{https://arxiv.org/abs/0706.2569}{{\ttfamily 0706.2569}}].

\bibitem{Gleisberg:2008ta}
T.~Gleisberg, S.~Hoeche, F.~Krauss, M.~Schonherr, S.~Schumann, F.~Siegert
  et~al., \emph{{Event generation with SHERPA 1.1}},
  \href{https://doi.org/10.1088/1126-6708/2009/02/007}{\emph{JHEP} {\bfseries
  02} (2009) 007} [\href{https://arxiv.org/abs/0811.4622}{{\ttfamily
  0811.4622}}].

\end{thebibliography}\endgroup

\end{document}